\documentclass[fleqn,usenatbib]{mnras}

\usepackage{newtxtext,newtxmath}

\usepackage[T1]{fontenc}

\DeclareRobustCommand{\VAN}[3]{#2}
\let\VANthebibliography\thebibliography
\def\thebibliography{\DeclareRobustCommand{\VAN}[3]{##3}\VANthebibliography}

\usepackage{graphicx}	%
\usepackage{amsmath}	%
\usepackage{xcolor}

\title[UV continuum slopes of galaxies at $z \geq 12$]{The ultraviolet continuum slopes of galaxies at $\mathbf{z \gtrsim 12}$ selected from ${\mathbf{0.6 \, deg}^2}$ of JWST imaging}

\author[C. Bondestam et al.]
{C. Bondestam,$^{1}$\thanks{\textbf{E-mail}: c.bondestam@ed.ac.uk}
D. J. McLeod $^{1}$,
F. Cullen$^{1}$,
R. J. McLure$^1$,
J. S. Dunlop$^1$,
S. Antonogiannaki$^1$,
\newauthor H.-H. Leung$^1$,
R. Begley$^2$, 
A. C. Carnall$^1$,
C. T. Donnan$^3$,
R. S. Ellis$^4$,
A. Fontana$^5$,
N. A. Grogin$^6$,
\newauthor I. J. B. Holst$^1$,
G. D. Illingworth$^7$,
A. M. Koekemoer$^6$,
F.-Y. Liu$^1$,
T. M. Stanton$^1$, 
S. D. Stevenson$^1$,
\newauthor  E. Taylor$^1$
\\
\\
\\
$^{1}$Institute for Astronomy, University of Edinburgh, Royal Observatory, Edinburgh EH9 3HJ, UK\\
$^{2}$Armagh Observatory and Planetarium, College Hill, Armagh, BT61 9DG, N. Ireland, UK\\
$^{3}$NSF's National Optical-Infrared Astronomy Research Laboratory, 950 N. Cherry Ave., Tucson, AZ 85719, USA\\
$^{4}$ Department of Physics and Astronomy, University College London, Gower Street, London WC1E 6BT, UK\\
$^{5}$INAF - Osservatorio Astronomico di Roma, Via Frascati 33, I-00078 Monteporzio Catone, Italy\\
$^{6}$Space Telescope Science Institute, 3700 San Martin Drive, Baltimore, MD 21218, USA\\
$^{7}$Department of Astronomy and Astrophysics, UCO/Lick Observatory, University of California, Santa Cruz, CA 95064, USA\\
}

\date{Accepted XXX. Received YYY; in original form ZZZ}

\pubyear{\the\year{}}

\begin{document}
\label{firstpage}
\pagerange{\pageref{firstpage}--\pageref{lastpage}}
\maketitle

\begin{abstract}
We present an analysis of the ultraviolet (UV) continuum slopes $(\beta)$ of $55$ galaxies at $z\gtrsim 12$, selected from a wide-area ($>0.6$ sq. deg) search across publicly available \textit{JWST}/NIRCam imaging. 
Our sample spans the absolute UV magnitude ($M_{\mathrm{UV}}$) range $-21.6 \leq M_{\mathrm{UV}} \leq -17.3$ and we find an inverse variance-weighted mean $\langle\beta\rangle = -2.57\pm0.04$ indicating that, \textit{on average}, the galaxy population is extremely blue at $z\gtrsim 12$.
The average $\beta$ at these redshifts is consistent with the intrinsic stellar plus nebular continuum limit in the absence of UV dust-attenuation.
Consistent with earlier \textit{JWST} works, our analysis suggests that the moderately dust-attenuated galaxies at $z \lesssim 10$ become essentially attenuation-free by $z \simeq 11$.
We find no evidence for a $\beta-M_{\mathrm{UV}}$ relation in our $z\gtrsim 12$ sample, indicating that the attenuation-free interpretation applies across all observed UV magnitudes.
By augmenting our sample with previous \textit{JWST} constraints down to $z=9$, and incorporating $M_{\mathrm{UV}}$-matched \textit{HST} studies down to $z=4$, we find a smooth evolution in $\beta(z)$ over a wide range in redshift. The average value of $\beta$ becomes progressively bluer at $z>4$, with $\mathrm{d}\beta/\mathrm{d}z \simeq -0.1$.
Finally, although we find that the typical galaxy at $z \gtrsim 12$ is consistent with the attenuation-free limit, the scatter in our individual $\beta$ measurements is consistent with the existence of a small population of dust-reddened galaxies with $\beta \gtrsim -2$.
\end{abstract}


\section{Introduction}\label{sec:introduction}

With the launch of \emph{JWST}, the redshift frontier of galaxy evolution has been pushed out to $z \simeq 15$ \citep[e.g.][]{Carniani2024, Naidu2026, donnan2026}, and the detection of reliable $z > 10 $ galaxy candidates is becoming commonplace thanks to the increasing area covered by deep \emph{JWST}/NIRCam imaging \citep{Donnan2024_UVLF, mcleod2024}. 
The large sample of known $z > 10$ galaxies is yielding crucial insights into the evolution of the population demographics via measurements of the ultraviolet (UV) luminosity function (LF) (see, e.g. \citealt{finkelstein2024,weibel2026,mcleod2026}).

Beyond the UVLF, deep \textit{JWST}/NIRCam photometry can also be used to determine the shape of the UV continuum in these ultra-high-redshift sources, characterized by the UV continuum slope, $\beta$, where $f_{\lambda} \propto \lambda^{\beta}$ for $\lambda_{\mathrm{rest}} \simeq 0.12 - 0.3 \, \mu \mathrm{m}$ \citep[e.g.][]{Calzetti1994, leitherer1999_sb99}.
The UV continuum shape encodes information about the intrinsic emission from massive stars (dependent on their age and metallicity), the nebular continuum emission from star-forming regions, and the degree of UV dust attenuation \citep[e.g.][]{Topping2022, Cullen2024}.

Stellar population models predict that the youngest, most metal-poor stellar populations set the so-called `blue limit' for the intrinsic stellar UV continuum slope at $\beta \simeq -3.2$ \citep[e.g.][]{stanway2016, Topping2022}.
Nebular continuum emission, dust attenuation, and increases in stellar age or metallicity all act to redden the UV slope (i.e. drive $\beta$ to less negative values).
At the highest redshifts, the effects of age and metallicity become subdominant \citep{Tacchella2022}, and the UV continuum is, to a good approximation, sensitive only to dust reddening and the strength of nebular continuum emission \citep[e.g.][]{Cullen2024, Katz2025}.
Many studies have now shown that for single-age stellar populations in the absence of dust, the blue limit when including the nebular continuum contribution is $\beta \simeq -2.6$ \citep{Topping2022, Cullen2024}\footnote{Accounting for a more realistic distribution of stellar population ages (i.e. relaxing the single-age assumption) can raise this attenuation-free limit to $\beta \simeq -2.4$ \citep[e.g.][]{Cullen2017}}.
Galaxies with $\beta$ at or bluer than this limit can be considered to be free of UV dust attenuation \citep{Cullen2024}.
Furthermore, galaxies with $\beta$ approaching the intrinsic stellar limit (i.e. $\beta < -2.8$) must necessarily also have a high ionising-photon escape fraction ($f_{\mathrm{esc}}$) to account for the absence of nebular continuum emission (e.g. \citealp{Chisholm2022, Saxena2026}; although see \citealp{katz2026} for another possible physical interpretation).

The picture is somewhat complicated by the existence of an extreme class of nebular-continuum-dominated galaxies that can exhibit red UV slopes ($\beta \simeq -2$) and mimic dust-reddened galaxies despite being dust-poor \citep[e.g.][]{cameron_2024_nebular}.
In these systems, the nebular continuum is so strong that the UV spectrum is dominated by two-photon emission from a highly-ionized interstellar medium.
Although the existence of such galaxies has been convincingly demonstrated \citep[e.g.][]{Katz2025}, their frequency among the overall galaxy population remains uncertain.
Fortunately, since these nebular-dominated galaxies exhibit red $\beta$ values, they are easily distinguishable from extremely dust-poor galaxies with $\beta \lesssim -2.6$ that are the primary focus of this work.

Prior to the launch of \emph{JWST}, significant effort was made to establish the evolution of typical $\beta$ values in the star-forming galaxy population from $z\simeq2$ to $z \simeq 8$.
At all redshifts, a correlation was found between $\beta$ and absolute UV magnitude ($M_{\mathrm{UV}}$) such that galaxies were observed to become younger and more dust-poor towards fainter magnitudes \citep[e.g.][]{rogers2014_colormag}.
Simultaneously, at fixed $M_{\mathrm{UV}}$, galaxies became increasingly blue towards higher redshifts, reaching average UV continuum slopes of $\langle \beta \rangle \simeq -2.1$ at $z=8$ for galaxies with $M_{\mathrm{UV}} \simeq -18$ \citep[e.g.][]{Finkelstein2012,Wilkins2012, Dunlop2013, bouwens2014_uvslopes}. 
Since the launch of \emph{JWST}, the evolution of the UV continuum slope has now been studied out to $z \simeq 12$ using large photometric samples \citep[e.g.][]{Topping2022, Topping2024,Cullen2023, Cullen2024, Morales2024, Austin2025, Austin2026,felicioni2026, Hainline2026}.
Spectroscopic samples are now also starting to be amassed beyond $z>10$ (including $10$ at $z>12$), making it possible to correlate $\beta$ to other physical properties \citep[e.g.][]{robertsborsani2024_prismstacks, Saxena2026, Pollock2026, robertsborsani2026_morestacking, Tang2026}. 
The consensus of these studies is that $\langle\beta\rangle$ continues to become bluer at fixed magnitude at $z \geq 8$, continuing the trend observed at lower redshifts. 
The evolution of the form of the $\beta-M_{\mathrm{UV}}$ relation with redshift is less certain, but with growing evidence for a flattening (i.e. no relation) at $z>10$ \citep{Saxena2026, Topping2024}.

One important result that has emerged from these \emph{JWST} studies is that by $z \simeq 11$ the average UV continuum slope of the observed galaxy population with $M_{\mathrm{UV}} \lesssim -18$ is consistent with the theoretical attenuation-free stellar plus nebular limit of $\beta = -2.6$ \citep[e.g.][]{Cullen2024, Topping2024}. 
This compares to typical values of $\langle\beta\rangle \simeq -2.2$ to $-2.3$ at $z \simeq 9$, just $\simeq 150 \, \mathrm{Myr}$ later \citep{Cullen2024, Topping2024}.

A clear transition to a bluer population on average at $z>10$ is also evident in spectroscopic samples, although recently several redder objects have also been reported \citep[e.g. ][]{mitsuhashi2026, Rodighiero2026}. 
The typical UV slopes of spectroscopic samples also tend to be slightly redder than those derived from photometry \cite[e.g.][]{Saxena2026, robertsborsani2026_morestacking}.

An obvious implication of these results is that at $z > 10$ a significant fraction of the galaxy population is unattenuated, which, for now, is also corroborated by the fact that there are still no confirmed dust continuum detections at $z > 9$ \citep[e.g.][]{Bakx2023, Kaasinen2023, popping2023_z12alma, Yoon2023, schouws2025_z14alma} whereas multiple detections have been achieved at $z = 8-9$ \citep[e.g.][]{laporte2017_dustz8, bakz2020_z8dust, bakx2025_z8dustcont}.
The lack of significant dust attenuation at these redshifts is obviously relevant to the debate concerning the number of UV-bright galaxies being observed at $z > 10$ with \emph{JWST}, and has implications for the number of ionising photons escaping into the intergalactic medium (IGM) at the beginning of the reionization era \citep[e.g.][]{munoz_2024}.

The physical explanation for a lack of dust attenuation in these ultra-blue high-redshift systems is still being debated, but some proposed ideas include dust being efficiently removed from early compact galaxies via supernova-driven winds \citep[e.g.][]{Ferrara2023, Fiore2023, Ziparo2023, Martinez-Gonzalez_2026}, an evolution in the physical properties of dust grains (specifically on average larger grains at high redshift; \citealp{ Narayanan_2025ApJ, Narayanan_2026_growthofdust}), a porous ISM \citep{Sommovigo2026} and dust destruction \citep{Esmerian2024, Garaldi2026}.
It has also been suggested that the average stellar mass of the population simply moves below the mass threshold at which dust attenuation is observed to become negligible at lower redshifts \citep{donnan2025_uvlfmodel}.
The debate remains unsettled, but it is nevertheless clear that constraints on the evolution of $\beta$ contain important information relating to the build up of dust and metals at early times.

Until now, most studies of UV continuum slopes have been limited to galaxies at $z < 12$  \citep[although see][]{Hainline2026, robertsborsani2026_morestacking}.
This is primarily due to the larger areas needed to compensate for the decreasing number density of sources with increasing redshift. 
However, the observable galaxy population is known to extend to at least $z \simeq 14.5$ \citep{Naidu2026}, and it is therefore important to investigate the evolution of $\beta$ out to this redshift frontier.
Fortunately, with the amount of deep \emph{JWST}/NIRCam imaging continually increasing, this analysis is now possible. 
In this paper, we examine the UV continuum slopes of $55$ galaxies at $z \gtrsim 12$ selected from $> 0.6 \, \mathrm{deg}^2$ of \emph{JWST} imaging.
The sample analysed here was first presented in \citet{mcleod2026} and used to constrain the evolution of the UV luminosity function.
Here, our aim is to use this sample to extend the analysis in \citet{Cullen2024} and investigate whether UV slopes continue to plateau at the $\beta = -2.6$ limit set by nebular physics.

The paper is structured as follows. 
In Section \ref{sec:catalogue_construction} we describe our selection procedure and the properties of the final sample.
In Section \ref{sec:results} we present our main results, specifically the evolution of $\beta$ with redshift and $M_{\mathrm{UV}}$.
Finally, we discuss the implications of our results and present comparison with other literature analyses in Section \ref{sec:discussion} and summarise our main results in Section \ref{sec:conclusions}.
Throughout we use the AB magnitude system \citep{oke1974,oke1983}, and assume a standard cosmological model with $H_0=70$\,km s$^{-1}$ Mpc$^{-1}$, $\Omega_m=0.3$ and $\Omega_{\Lambda}=0.7$.

\section{Sample Selection and Properties}\label{sec:catalogue_construction}

The present study focuses on a sample of $z\gtrsim 12$ galaxies that was first presented in \citet{mcleod2026}. 
These galaxies were selected over an extensive, wide-area search, encompassing $>0.6 \, \mathrm{deg}^2$ and over $\simeq 150$ sight-lines, enabling us to produce a large and robust $z \gtrsim 12$ galaxy sample while mitigating the effects of cosmic variance. 
We refer interested readers to \citet{mcleod2026} for a complete description of the catalogue construction and high-redshift galaxy selection, but give a summary of the most relevant details below.

\subsection{The $\mathbf{z\gtrsim 12}$ sample}\label{sec:catalogue_construction_z_grt_12}

\begin{figure}
    \centering
    \includegraphics[width=\linewidth]{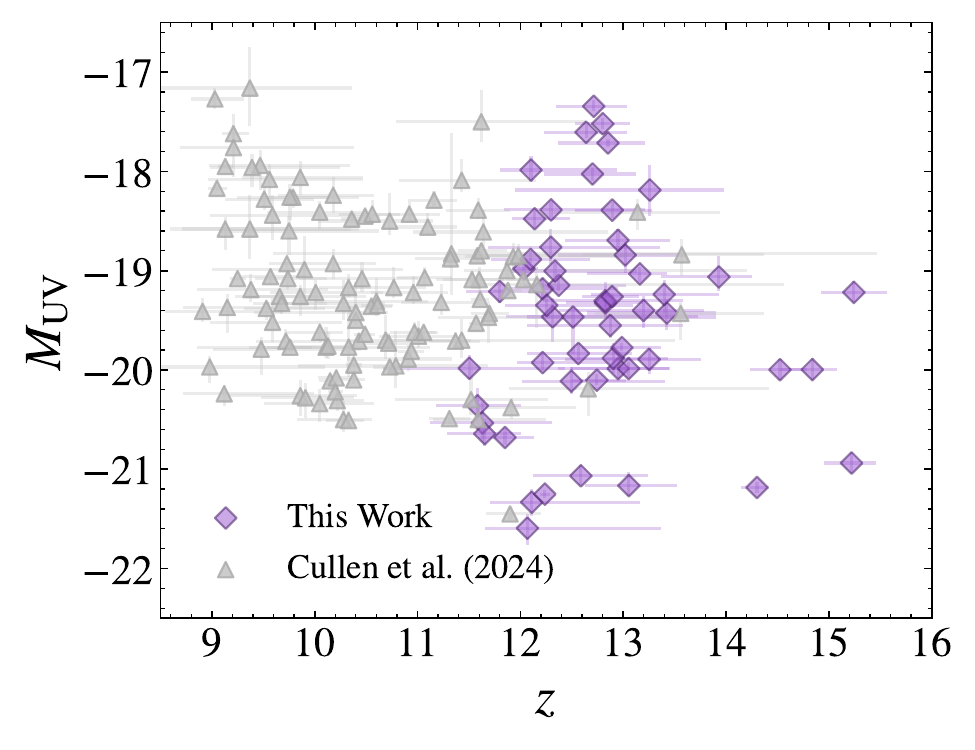}
    \caption{The distribution of UV absolute magnitude $M_\mathrm{UV}$ and photometric redshift $z$ for our high redshift galaxy sample.
    The purple diamonds show our new sample of $55$ galaxy candidates at $z \gtrsim 12$ selected from $>0.6~\mathrm{deg}^2$ of deep \emph{JWST}/NIRCam imaging (see \citealp{mcleod2026} for a full description of this sample). 
    The grey triangles show a complementary sample of galaxies at $9 < z < 12$ taken from our previous analysis \citep{Cullen2024}. 
    Our new $z\gtrsim 12$ sample spans the redshift range $11.5 \leq z \leq 15.2$ and absolute UV magnitude range $-21.6 \leq M_\mathrm{UV} \leq -17.3$.}
    \label{fig:MUV_vs_z}
\end{figure}

Across each of the fields, multi-wavelength photometry catalogues were produced with \textsc{Source Extractor} \citep{bertin1996} in dual-image mode. 
These catalogues were constructed using three different detection images: $F200W$, $F277W$ and $F356W$. 
The initial catalogues were culled such that we only considered $F150W$ dropout objects--i.e. with a signal-to-noise ratio ($\mathrm{SNR}$) of  $\mathrm{SNR} \leq2$ for $F150W$ and all bluer filters--which also had $\mathrm{SNR} \geq 8$ in at least one of either the $F200W$ or $F277W$ images. 
As an additional step to exclude obvious artefacts, we also required $\mathrm{SNR} \geq 3$ in at least one other band redward of the Lyman break. 
Aperture photometry was performed in $0.2^{\prime\prime}-$diameter apertures on images that were homogenized to match the resolution of $F444W$. 
The photometric uncertainties were calculated source-by-source as an aperture-to-aperture root mean square of the nearest $200$ blank-sky apertures, as determined via a dilated segmentation map (see, e.g., \citealt{begley2025} for a full description of this method).

Photometric redshifts were estimated with the spectral energy distribution (SED) fitting code $\textsc{LePhare}$ \citep{arnouts2011}, using the BC03 template library \citep{bc2003}, assuming a \citet{chabrier2003} initial mass function and a \citet{Calzetti1994} attenuation law. 
Objects with $z_{\rm phot}\geq11.5$ and a $\Delta\chi^{2}\geq4$ between the primary high-redshift solution and the secondary low-redshift solution were retained, and subject to rigorous visual inspection to remove any remaining artefacts or lower redshift interlopers.
The result of this process was a final sample of $55$ galaxies, within the redshift range $11.5 \leq z_\mathrm{phot} \leq 15.2$ and absolute UV magnitude range $-21.6 \leq M_\mathrm{UV} \leq -17.3$ (Fig. \ref{fig:MUV_vs_z}; \citealp{mcleod2026}). 

\subsection{Ancillary $\mathbf{9 < z < 12}$ sample}

In addition to our new $z\gtrsim 12$ sample, we also make use of the complementary $z\geq9$ galaxy sample presented in our previous work that was selected following a very similar methodology \citep{Cullen2024}. 
This sample was selected from a search for $z>9$ galaxies across $\simeq320$ sq. arcmin and fifteen independent datasets and was an extension of the $z>9$ galaxy sample first presented in \citet{mcleod2024}. 
Photometry was measured in $0.35^{\prime\prime}$-diameter apertures on imaging that was PSF-homogenized to match the resolution of $F444W$. 
Similar to the present study, we adopted a stringent $\mathrm{SNR}\geq8$ criterion in at least one of the $F150W$, $F200W$ or $F277W$ images, with a requirement $\mathrm{SNR}\leq2$ at bluer wavelengths. 
The SED fitting setup with \textsc{LePhare} and candidate selection procedure was identical to the one described above.
This sample consists of $121$ sources selected with $z_{\mathrm{phot}}\geq9.0$.
Barring a small number of $z > 12$ candidates (see Fig. \ref{fig:MUV_vs_z}), this sample is practically confined to the redshift range $9 < z < 12$ and absolute UV magnitude range ${-20.5<M_{\mathrm{UV}}<-17.0}$. 
We refer the interested reader to \citet{Cullen2024} for a full description of the data set and sample selection.

We remove two galaxies from the \citet{Cullen2024} sample that, when including new data, are no longer consistent with being $z > 9$ galaxy candidates (IDs JADES-1058823 and NGDEEP-1003756 from table 1 of \citealp{Cullen2024}).
We also remove a further four galaxies that are duplicated in our new selection (IDs GLASS-9974, GLASS-28072, CEERS-37967, and JADES-9996) from table 1 of \citealp{Cullen2024}.
This leaves a total of $170$ sources at $z \geq 9$; the full distribution in $z$ and $M_{\mathrm{UV}}$ is shown in Fig. \ref{fig:MUV_vs_z}.
It can be seen that with these new $z\gtrsim 12$ candidates, we can naturally extend our previous analysis of UV continuum slopes at $9 < z < 12$ \citep{Cullen2024}.
Both samples span a very similar range in $M_{\mathrm{UV}}$, although the factor $\simeq 7$ increase in area means that we have been able to find a large number of bright candidates ($M_{\mathrm{UV}} \lesssim -21$) in the new $z \gtrsim 12$ search.

\begin{figure*}
    \centering
    \includegraphics[width=0.95\linewidth]{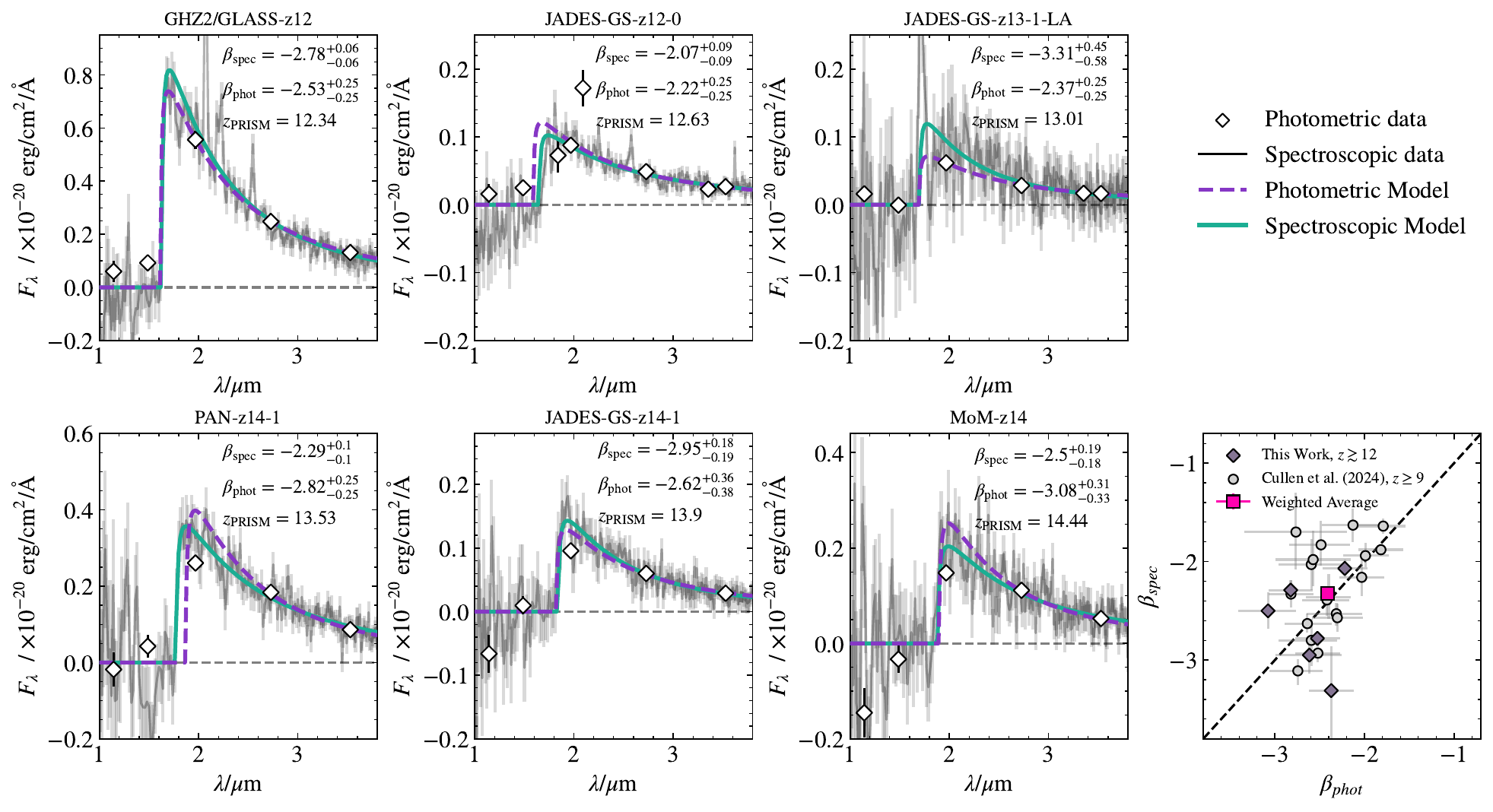}
    \caption{The first six panels show examples of our photometric ($\beta_{\mathrm{phot}}$) and spectroscopic ($\beta_{\mathrm{spec}}$) UV continuum model fits to the six galaxies in our sample with available \emph{JWST}/NIRSpec PRISM spectra at $z>12$ (see Section \ref{sec:uvslope_fitting_spectroscopy}).
    The black lines show the \emph{JWST}/NIRSpec PRISM spectra with the grey shading illustrating the $1\sigma$ uncertainties on the flux densities.
    The white diamonds show the photometry for each object obtained from \emph{JWST}/NIRCam imaging data.
    Our best-fitting models obtained when fitting to the spectroscopic data are shown by the solid teal lines, with the corresponding best-fitting models fitted to the photometric data shown as dashed purple lines.
    The best-fitting values of $\beta_{\mathrm{spec}}$ and $\beta_{\mathrm{phot}}$ are given in the legend of each panel.
    The final panel (bottom-right) shows a plot of ${\beta_{\mathrm{spec}} \, \mathrm{versus} \, \beta_{\mathrm{phot}}}$ with the $z\gtrsim12$ galaxies shown as dark-grey diamonds. 
    The light grey circles show an ancillary sample of $z\ge9$ galaxies from \citet{Cullen2024}. 
    The pink square is the inverse variance-weighted average across all of these objects. 
    Overall, we find good agreement between the averages with $\langle \beta_{\mathrm{spec}}\rangle = -2.32\pm 0.05$ and $\langle \beta_{\mathrm{phot}}\rangle = -2.41\pm 0.08$.}
    \label{fig:beta_spec_vs_beta_phot_comparison}
\end{figure*}

\subsection{Measuring the UV continuum slope}\label{sec:uvslope_fitting_photometry}

To fit the UV continuum slopes for our sample we follow the method outlined in \citet{Cullen2024} with minor adjustments.
Briefly, the UV continuum is modelled as a power law following $f_{\lambda} \propto \lambda^\beta$ for $\lambda_{\mathrm{rest}} \leq 3000$~\AA~\citep[][]{Calzetti1994}. 
The effect of the intergalactic medium (IGM) is accounted for by setting all $f_{\lambda}$ at $ \lambda_{\mathrm{rest}} \leq 1216$~\AA~to zero and incorporating the Ly$\alpha$ damping wing using equation 2 of \citet{Miralda_escude1998}. 
Given the extreme redshifts of our objects ($z\gtrsim12$), we choose to fix the fraction of neutral hydrogen in the surrounding IGM ($x_{\mathrm{HI}}$) to $1$.

The model therefore has two free parameters: the UV continuum slope ($\beta$) and the redshift ($z$).
To fit this model to the photometric data we use the nested sampling algorithm \texttt{dynesty} \citep[][]{Speagle2020, koposov2026}.
For the UV continuum slope, we assume a flat prior in the range ${-10 \leq \beta \leq 10}$.
For the redshift prior, we sample from the individual redshift probability distributions, $p(z)$, returned from fitting to the full SED.
Since \textsc{LePhare} does not include a Ly$\alpha$ damping prescription, to obtain a more accurate $p(z)$ we refit the SEDs using a custom fitting code, described in \citet{mclure2011}, that has been modified to account for Ly$\alpha$ damping.
The best-fitting parameter values returned by the \texttt{dynesty} fits are used throughout our analysis unless otherwise specified. 
The reported uncertainties are estimated using the 16th and 84th percentiles of the posterior probability distribution for each variable. 
We also adopt a minimum error of $\sigma_\beta = 0.25$ for the UV continuum slope, following the discussion in \citet{Cullen2024}. 
Some examples of fits to photometric data can be found in Fig. \ref{fig:beta_spec_vs_beta_phot_comparison} (purple dashed lines).

It is worth noting that we do not fix the redshift to the photometric redshift from the full SED fitting; the redshift values and uncertainties used throughout this paper are the ones obtained with our own model fits.
We do this in order to marginalise over the redshift uncertainties consistent with our assumed model.
However, we have verified that fixing the redshift to the best-fitting SED value does not change our results (see Fig. \ref{fig:fixed_z_plot}).
The median difference in $\beta$ between our method (allowing $z$ to vary) and a fixed-$z$ approach is $\Delta\beta = -0.06$, with the median redshift difference being $\Delta z = z_\mathrm{\texttt{dynesty}} - z_\mathrm{SED} = 0.06$.

Finally, although a number of sources in our sample have confirmed spectroscopic redshifts (see Section \ref{sec:uvslope_fitting_spectroscopy}) we do not fix the redshift for these sources either, so as not to bias our weighted means to galaxies with the most accurate constraints.
As we will discuss in Section \ref{sec:uvslope_fitting_spectroscopy} below, we show via a comparison between the photometric and spectroscopic $\beta$ measurements that, although on a source-by-source basis the photometric estimates do not recover the correct UV slope, the weighted mean of a sample of individual measurements, $\langle \beta_{\mathrm{phot}}\rangle$, should be fully consistent with $\langle \beta_{\mathrm{spec}}\rangle$.

\subsubsection{Absolute UV Magnitudes}

To determine absolute UV magnitudes and their corresponding uncertainties, we sample models from the full posterior distribution and integrate each through a top-hat filter centred on $ \lambda_{\mathrm{rest}} =  1500$~\AA~ with a width of $100$~\AA . 
The uncertainties are estimated using the 16th and 84th percentiles of the resulting $M_{\mathrm{UV}}$ distribution. The $M_{\mathrm{UV}}$ values are corrected to total by scaling to the \textsc{FLUX\_AUTO} measurement from \textsc{Source Extractor} with an additional 10 per cent correction, to account for light outside the Kron aperture \citep{Kron1980}.
For any object in a lensed field, we further correct by the magnification factor ($\mu$) using the relevant lensing map as referenced in Table \ref{tab:appendix_all_sources}. 
Only a small number of our objects are lensed: five have a modest magnification $(\mu<2)$ and one is moderately lensed $(\mu \simeq 3.2)$.

\subsection{Spectroscopic control sample}\label{sec:uvslope_fitting_spectroscopy}

The results presented in this study are based on an analysis of broadband and medium-band photometry of photometrically-selected galaxy candidates at $z>9$.
The main advantage of this approach is that it enables the analysis of a large galaxy sample with a well-defined selection function.
However, one obvious drawback of a photometric analysis is that individual UV continuum slope measurements ($\beta_\mathrm{phot}$) are much less accurate than can be achieved with spectroscopy ($\beta_\mathrm{spec}$) due to the significantly decreased spectral resolution of photometric data.
Fortunately, a small subset of our photometric sample has \emph{JWST}/NIRSpec PRISM data  available, which allows us to assess the accuracy of the photometric measurements.
Specifically, we can check that, despite the large uncertainties of individual $\beta_\mathrm{phot}$ measurements, they can still accurately recover the average $\beta$ values of a given galaxy sample.

The basic methodology for measuring $\beta_\mathrm{spec}$ from \emph{JWST}/NIRSpec PRISM is the same as for measuring $\beta_\mathrm{phot}$.
For fitting to the spectroscopic data the model is smoothed and resampled to match the resolution and wavelength sampling of the NIRSpec/PRISM data and compared on a pixel-by-pixel basis \citep[see][]{Cullen2024}.
To minimize the effect of emission lines on our $\beta_\mathrm{spec}$ measurement, we mask the regions contaminated by strong emission or absorption line features using the \citet{Calzetti1994} windows.
For one galaxy (JADES-GS-z13-1-LA; \citealp{witstok_2025_z13_lya}) we additionally mask out the strong $\mathrm{Ly}\alpha$ line that is present in its spectrum.

Out of the $55$ objects in our $z\gtrsim 12$ sample, $6$ have available \emph{JWST} NIRSpec/PRISM spectra:  GHZ2/GLASS-z12 \citep{Castellano2024}, JADES-GS-z12-0 \citep[][]{Curtis-Lake_2023}, JADES-GS-z13-1-LA \citep{witstok_2025_z13_lya}, PAN-z14-1 \citep{donnan2026}, JADES-GS-z14-1 \citep{Carniani2024}, and MoM-z14 \citep[][]{Naidu2026}. 
All of these, except PAN-z14-1, are publicly available and fully reduced on the \textit{DAWN JWST Archive} \citep[DJA; see][for descriptions of the reduction process]{Heintz2024_DJA,deGraaff_DJA}.
PAN-z14-1 was reduced as described in \citet{Arrabal_Haro2023_ceers_spectra} with modifications as described in \citet{donnan2026}. 
The model fits to these six objects are shown in Fig. \ref{fig:beta_spec_vs_beta_phot_comparison} as teal lines.

We note that there are two other known $z_{\rm spec}>12$ galaxies that are not in our sample: JADES-GS-z14-0 \citep[][]{Carniani2024} and JADES-GS-z13-0 \citep[][]{Curtis-Lake_2023}. 
The former is not selected due to its close proximity to a nearby low redshift foreground galaxy, while the latter is too faint to pass our photometric selection criteria, specifically the requirement of $\rm SNR\geq8$ in at least one of either the $F200W$ or $F277W$ images. 

To expand the $\beta_\mathrm{phot}-\beta_\mathrm{spec}$ comparison we also incorporate objects from the ancillary \citet{Cullen2024} sample. 
Of the $115$ objects in this sample, $22$ have \emph{JWST} NIRSpec/PRISM spectra publicly available on the DJA. 
For three of these spectra, we were unable to obtain a successful fit and they were removed from the sample.
In total, therefore, we have a sample of $\mathrm{N}=25/170$ galaxies with measurements of both $\beta_\mathrm{phot}$ and $\beta_\mathrm{spec}$, spanning the full redshift range of our analysis ($9 < z < 16$).

Fig. \ref{fig:beta_spec_vs_beta_phot_comparison} shows our model fits to both the photometric data (dashed purple line) and spectroscopic data (solid teal line) for the $6$ objects with \emph{JWST}/NIRSpec PRISM spectra at $z>12$. 
The bottom right-hand panel of Fig. \ref{fig:beta_spec_vs_beta_phot_comparison} shows the comparison between $\beta_{\mathrm{spec}}$ and $\beta_{\mathrm{phot}}$ for all $25$ galaxies with spectroscopic measurements, with the pink square showing the inverse variance-weighted average of all of the individual measurements. 
Overall, we find good agreement between $\beta_{\mathrm{spec}}$ and $\beta_{\mathrm{phot}}$ with 54 per cent of the values being consistent within the $1\sigma$ uncertainties, and all values within $2 \sigma$ uncertainties.

The two weighted averages are also consistent at the $\simeq 1 \sigma$ level, with $\langle \beta_{\mathrm{spec}}\rangle = -2.32\pm0.05$ and $\langle \beta_{\mathrm{phot}}\rangle = -2.41\pm 0.08$, indicating that our photometric model is a good predictor of the UV continuum slope, especially when averaging over a large sample size.
If we impose a minimum uncertainty floor of $\sigma_{\beta}=0.1$ on the spectroscopic measurements to avoid the average being strongly biased by higher SNR spectra, the weighted average moves closer to the photometric value, $\langle \beta_{\mathrm{spec}}\rangle = -2.35\pm0.03$.
In our previous work \citep{Cullen2024} we also found an offset of $0.06$ between $\langle \beta_\mathrm{spec} \rangle$ and $\langle \beta_\mathrm{phot} \rangle$ in the same direction (i.e. $\langle \beta_\mathrm{phot} \rangle < \langle \beta_\mathrm{spec} \rangle$).
Taken together, these comparisons suggest that our $\beta_{\mathrm{phot}}$ averages are likely to be biased slightly blue.
We will return to this issue of a potential bias in our $\beta_{\mathrm{phot}}$ estimates later (Section \ref{sec:simulation_section}).

\begin{figure*}
    \centering
    \includegraphics[width=0.8\linewidth]{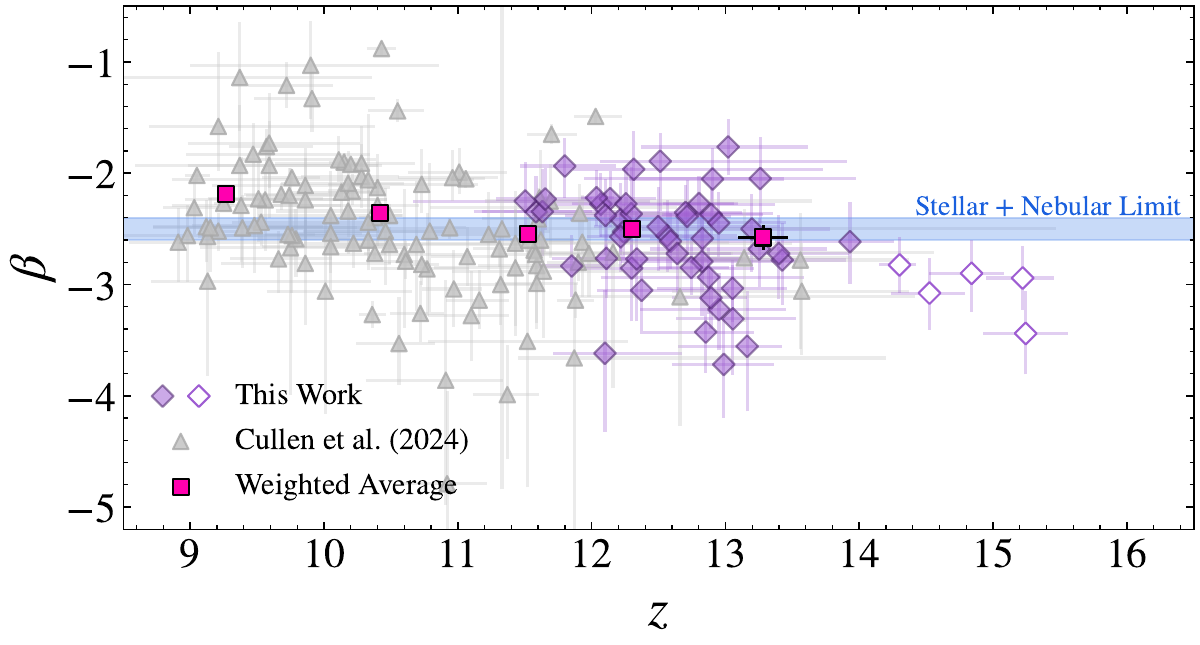}
    \caption{The UV continuum slope $(\beta)$ plotted against redshift $(z)$ for photometrically-selected galaxy candidates at $z\geq9$.
    The purple diamonds show our new $z\gtrsim 12$ sample, with the unfilled points indicating $z>14$ galaxies with more tentative measurements that are not included when fitting $\beta-z$ relations. 
    The grey triangles show the ancillary $z\ge9$ sample from \citet{Cullen2024}. 
    The region between $\beta = -2.4$ to $\beta=-2.6$ has been highlighted with a light blue shading to show the theoretical range of $\beta$ that represents the intrinsic stellar plus nebular continuum limit. 
    At these values of $\beta$ it is likely that galaxies are free of dust attenuation in the rest-frame UV (see text for discussion). 
    The inverse variance-weighted averages in bins of redshifts are shown as pink squares and reach values of $\langle\beta\rangle\simeq-2.6$ at $z\gtrsim11$.
    These results indicate that --- although individual reddened galaxies can exist --- the galaxy population on average is consistent with the intrinsic limit and suffers negligible UV dust attenuation.}
    \label{fig:beta_vs_z}
\end{figure*}

\section{Results}\label{sec:results}

In this section we first discuss the range of measured $\beta$ values within our new $z \gtrsim 12$ sample, and assess evidence for exotic ultra-blue individual sources. 
We then investigate the redshift evolution of $\langle \beta \rangle$ both within our sample as well as more broadly across the redshift range $z\simeq 15$ to $z\simeq4$ by incorporating the results from previous works.
Finally, we discuss evidence for a $\beta-M_{\mathrm{UV}}$ relation within our $z \gtrsim 12$ sample.

\subsection{UV continuum slopes at $\mathbf{z \gtrsim 12}$}
\label{sec:beta_high_redshift}

In Fig. \ref{fig:beta_vs_z} we show $\beta_{\mathrm{phot}}$ as a function of redshift ($z$) for our primary $z \gtrsim 12$ sample (purple diamonds) and $9 < z < 12$ sample (grey triangles; \citealp{Cullen2024}).
The blue shaded region spans $-2.6 \leq \beta \leq -2.4$, with $\beta \simeq -2.6$ representing the minimum \emph{intrinsic} $\beta$ value of a stellar plus nebular continuum SED with $f_\mathrm{esc} = 0$.
However, it is worth noting that this is a minimum value across all stellar population models, stellar ages and metallicities.
For reasonable assumptions regarding the stellar populations of $z>9$ star-forming galaxies, the true intrinsic UV slopes are likely between $\beta \simeq -2.6$ and $\beta \simeq -2.4$ (e.g. see fig. 2 of \citet{Topping2022} or fig. 9 of \citealp{Cullen2024} and the corresponding discussions).
UV slopes at or below this intrinsic limit must necessarily be free of dust attenuation.
The primary result of \citet{Cullen2024} was that by $z \simeq 11$ the \emph{typical} UV slope of the galaxy population (with $-20 \lesssim M_{\mathrm{UV}} \lesssim -17$) had reached this blue-limit, with an inverse variance-weighted mean of $\langle \beta \rangle = -2.59 \pm 0.06$ at $z = 11.5$. 
The primary motivation of this work is to determine whether this result holds with the addition of our new $z \gtrsim 12$ sample.

Looking first at the individual $\beta$ measurements, it is clear that we observe a significant scatter, with values ranging between $\beta = -1.76$ and $\beta=-3.72$.
This is in large part due to the significant uncertainties associated with individual photometric estimates, with the median uncertainties of the sample being $\langle\sigma_{\beta, \rm upper}\rangle = 0.31$ and $\langle\sigma_{\beta, \rm lower}\rangle = 0.34$.
In fact, taking the large typical uncertainties at face value, it is likely that the sample is drawn from a population with a relatively narrow range of $\beta$.
However, larger spectroscopic samples will be needed to definitively confirm the true distribution of $\beta$ values at these redshifts.

Of the 55 galaxies in the primary sample, $11$ have extremely blue UV slopes below or at the canonical `stellar-limit' of $\beta = -3.2$, with the bluest observed value being $\beta = -3.72^{+0.40}_{-0.48}$.
Definitive confirmation of sources with $\beta < -3$ would be very interesting, as these are likely extremely young, perhaps even metal-free, stellar populations with extremely high escape fractions \citep[e.g.][]{schaerer2002_pop3}.
Their existence would have important implications for our understanding of early star-formation and the process of reionization.
However, from our ultra-blue sub-sample, we find that $6/11$ are consistent within $1\sigma$ of $\beta = -3.00$ and all are within $2\sigma$ of this value. 
Moreover, all eleven sources are also consistent within $3\sigma$ of the $\beta = -2.6$ attenuation-free limit. 
In other words, although they almost certainly have very steep UV slopes, none can be considered \textit{robustly} $\beta \lesssim -3$ sources.

Despite this, the fact that $20$ per cent of our sample have measured values of $\beta \leq -3$ is one clear indication that the galaxy population at $z\gtrsim 12$ must be, on average, very blue. 
Indeed, we find that the inverse variance-weighted average for our full $z\gtrsim 12$ sample is $\langle \beta \rangle =-2.57\pm0.04$.
This is consistent with the intrinsic $\beta$ limit, indicating that the typical galaxy at $z\gtrsim 12$ is essentially free of UV dust attenuation.
In the following sections, we consider the evolution of $\langle \beta \rangle$ with redshift and $M_{\mathrm{UV}}$ in more detail.

\subsection{The redshift evolution of the UV continuum slope}\label{sec:beta_redshift_evolution}

\begin{table}
    \centering
        \caption{The inverse variance-weighted redshift $\langle z \rangle $ and UV continuum slope $\langle \beta \rangle $ for our combined sample of 170 galaxies ($z>9$ galaxies from \citet{Cullen2024} and $z \gtrsim12$ galaxies from \citet{mcleod2026}). The five bins are presented in Fig. \ref{fig:beta_vs_z} and Fig. \ref{fig:beta_vs_z_weighted_average}.}
    \begin{tabular}{|l|c|c|c|}
    \hline
       Redshift Range & Number & $\langle z \rangle$& $\langle\beta\rangle$ \\
       \hline
         $z<10$ &37& $\phantom{0}9.27\pm0.04$  & $-2.19\pm0.06$\\
         $10\leq z < 11$ & 41 & $10.42\pm0.03$ & $-2.36\pm0.05$ \\
         $11\leq z < 12$ & 37 & $11.52\pm0.04$  & $-2.55\pm0.05$\\
         $12\leq z < 13$ & 37 & $12.30\pm0.04$ & $-2.50\pm0.05$ \\
         $13 \leq z <14$ & 13& $13.28 \pm0.19$ & $-2.58\pm0.11$ \\
        \hline
    \end{tabular}

    \label{tab:weighted_averages_beta}
\end{table}

An evolution towards bluer $\beta$ is expected at higher redshifts as galaxies will have progressively younger stellar populations, lower metallicities and consequently lower dust attenuations \citep[e.g.][]{Robertson2010, Tacchella2022}. 
One important result from early \textit{JWST} studies has been that by $z\simeq11$ the average value of $\beta$ has reached the intrinsic limit of $\beta \simeq -2.6$ \citep[e.g.][]{Cullen2024, Topping2024}. 
Whether or not this remains true for the galaxy population at $z \gtrsim 12$ remains an open question.
Although current spectroscopic samples at $z>12$ suggest the galaxy population does remain ultra-blue \citep[e.g.,][]{Pollock2026, robertsborsani2026_morestacking, Tang2026}, this has yet to be tested with larger photometric samples.

Fig. \ref{fig:beta_vs_z} shows the $\beta-z$ relation combining our new $z\gtrsim 12$ sample (purple diamonds) with the \citet{Cullen2024} sample at $9 < z < 12$ (grey triangles).
Weighted averages in five bins of redshift are shown by the pink squares, and the corresponding values are given in Table \ref{tab:weighted_averages_beta}.
In the redshift range $9 < z < 12$ we reproduce the result reported in \citet{Cullen2024} --- despite the minor changes to the sample in this redshift range.
The average UV continuum slope evolves from ${\langle \beta \rangle = -2.19 \pm 0.06}$ at $z \simeq 9.3$ to ${\langle \beta \rangle = -2.55 \pm 0.05}$ at $z \simeq 11.5$.
As discussed in \citet{Cullen2024}, the straightforward interpretation of this result is that, while the typical galaxy at $z<10$ is consistent with moderate amounts of UV attenuation (i.e. having $\beta > -2.4$), by $z\simeq11$ the population average is consistent with the expected intrinsic attenuation-free value of $\beta$.

In the additional redshift bins that we can now access with our $z\gtrsim 12$ sample we find ${\langle \beta \rangle = -2.50\pm0.05}$ at $z \simeq 12.3$ and ${\langle \beta \rangle = -2.58\pm0.11}$ at $z\simeq 13.3$, suggesting that the general trend is for the galaxy population to remain consistent with the intrinsic limit at $z > 12$.
At $z \geq 14$, our sample contains very few galaxies ($\mathrm{N}=5$) and we therefore do not consider the sample statistics sufficient to draw robust conclusions. Given the lack of robust photometric candidates at $z>14$ to date, we treat these sources as tentative, presenting them as unfilled datapoints in Fig. \ref{fig:beta_vs_z}.
However, it is interesting to note that all of the sources at these redshifts fall below $\beta=-2.6$, with the weighted average being ${\langle \beta \rangle = -2.99 \pm 0.15})$.
There are two spectroscopically-confirmed $z > 14$ galaxies of which one, MoM-z14 (or PRIMER-COS-2332181 in Table \ref{tab:appendix_all_sources}; \citealp{Naidu2026}), makes it into our sample selection (see Section \ref{sec:uvslope_fitting_spectroscopy} for details on why the galaxy discovered in \citealp{Carniani2024} is not included).
We fit a spectroscopic UV slope of $\beta_{\mathrm{spec}}=-2.50^{+0.19}_{-0.18}$ for MoM-z14 , which is redder than our photometric constraint of $\beta_{\mathrm{phot}}=-3.08^{+0.31}_{-0.38}$ (see Table \ref{tab:appendix_all_sources}).
The other confirmed $z > 14$ galaxy has a spectroscopic UV slope of $\beta_{\mathrm{spec}}=-2.20\pm0.07$, as reported in \citet{Carniani2024}.
It is therefore possible that a stronger bias affects photometric constraints at the highest redshifts.
Indeed, our $\beta$ recovery simulations do show an increasing blue bias at $z > 14$ (see Section \ref{sec:simulation_section}). 
Ultimately however, more $z > 14$ sources (including robust photometric candidates) are needed to make progress here.
For the analysis below, we do not include the poorly-constrained $z > 14$ bin in our $\beta-z$ fitting analysis, and note that excluding these sources does not significantly change our $z\gtrsim12$  weighted average with the value going from $-2.57\pm0.04$ to $-2.52\pm0.05$.

\begin{figure*}
    \centering
    \includegraphics[width=\linewidth]{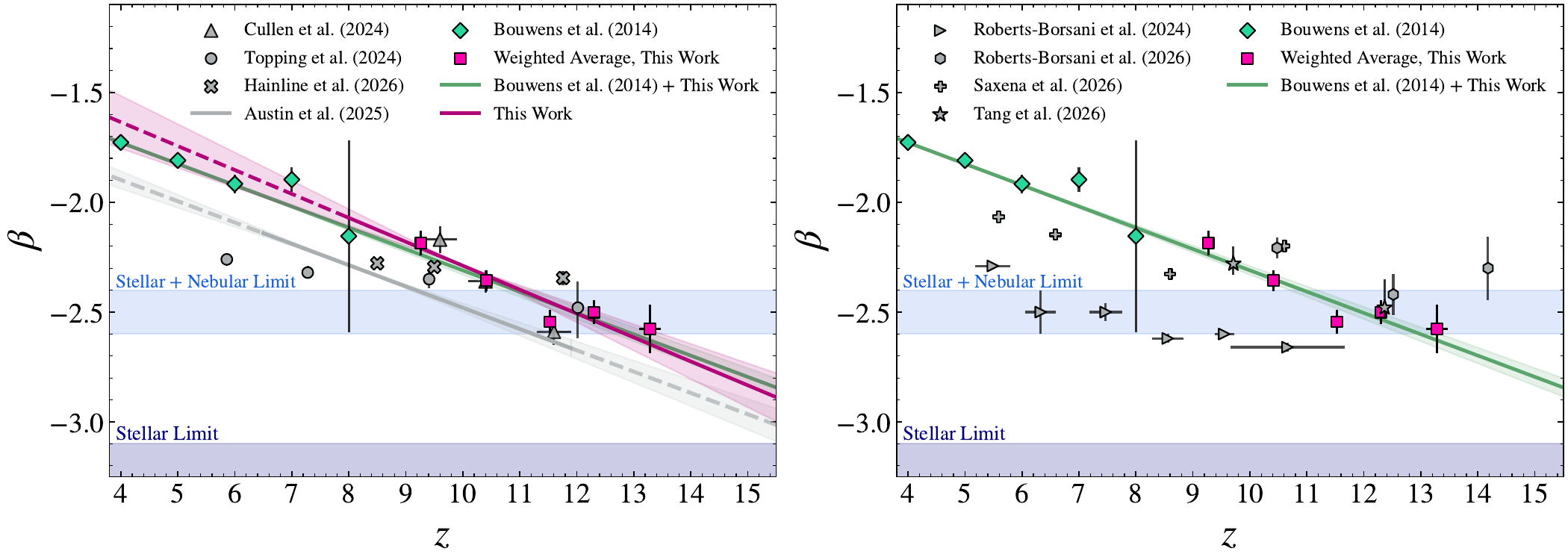}
    \caption{The left-hand panel shows a $\beta-z$ comparison between the results presented here and other photometric analyses from the literature \citep[][]{Cullen2024, Topping2024, Austin2025, Hainline2026}. 
    The pink line, which has a gradient of $\mathrm{d}\beta/\mathrm{d}z = -0.11\pm0.02$ is a linear fit to our inverse variance-weighted averages at $z\ge9$ (pink squares), with the shaded region showing the $1\sigma$ uncertainty bounds. 
    We combine our measurements with an $M_{\mathrm{UV}}$-matched sample from \citet{bouwens2014_uvslopes} that extends the relation down to $z=4$ (green diamonds).
    It is clear that these data closely follow the extrapolation of our fit at $z\ge9$.
    A combined fit to our measurements and the \citet{bouwens2014_uvslopes} points yields a slope of $\mathrm{d}\beta/\mathrm{d}z = -0.097\pm0.004$, which is consistent within the $1\sigma$ uncertainty.
    The right-hand panel shows the same $\beta-z$ plane, but now comparing our data (and \citealp{bouwens2014_uvslopes}) to the spectroscopic constraints derived from stacked \emph{JWST}/NIRSpec PRISM data presented in \citet{robertsborsani2024_prismstacks},  \citet{robertsborsani2026_morestacking}, \citet{Saxena2026}, and \citet{Tang2026}.
    Although significant variance exists among the current spectroscopic measurements, the level of agreement at $z>10$ is generally good.}
    \label{fig:beta_vs_z_weighted_average}
\end{figure*}

\subsubsection{The transition to an attenuation free population}

Given the indication in Fig. \ref{fig:beta_vs_z} that the typical UV continuum slope reaches the intrinsic limit at $z \gtrsim 10$, we consider the redshift above which the galaxy population can be considered to be --- on average --- UV attenuation-free.
To do this, we follow a similar approach to \citet{Cullen2024}, who found based on their $9 < z < 12$ sample that a transition to $\langle \beta \rangle \simeq -2.6$ occurred at $z > 10.5$.
We fit two toy models to the individual $\beta$ measurements across both samples: (i) a step function and (ii) a piecewise function consisting of a linear relation (with slope $m$) transitioning to a constant value. 
The step function has the following form:
\begin{equation}
    \beta_\mathrm{step}(z) = \begin{cases}
            \beta_0  & \text{if } z < z_{t,\mathrm{step}} \\
             \beta_1  & \text{if } z \ge z_{t,\mathrm{step}}
       \end{cases},
\end{equation}
and the piecewise function takes the form: 
\begin{equation}
    \beta_\mathrm{pw}(z) = \begin{cases}
             mz+\beta_0  & \text{if } z < z_{t,\mathrm{pw}} \\
             \beta_1  & \text{if } z \ge z_{t,\mathrm{pw}}.        \end{cases}
\end{equation}
Although neither model is strongly physically motivated, the fits nevertheless return a useful indication of a transition redshift ($z_t$) above which the average galaxy population reaches its bluest value ($\beta_1$). 
The two transition redshifts we find are $z_{t,\mathrm{step}} = 10.56^{+0.03}_{-0.05}$ and $z_{t, \mathrm{pw}}=10.56^{+0.05}_{-0.02}$, respectively.
At redshifts above this threshold, the plateau value is $\beta_1 = -2.54^{+0.04}_{-0.03}$.
In both cases these are consistent with findings in \citet{Cullen2024} and again highlight the fact that the average UV continuum slopes at $z \gtrsim 11$ become consistent with the intrinsic stellar plus nebular limit.
From these fits, we can say that $z \simeq 11$ roughly marks a transition epoch above which the UV SED of the typical galaxy is not experiencing dust attenuation.
We note that this is a conservative upper limit on the redshift, since $\beta=-2.6$ is the hard limit for an attenuation free stellar population but any galaxy with $\beta \leq -2.4$ could plausibly be unattenuated.

\subsubsection{Extending the $\beta-z$ relation to $z=4$}\label{sec:comparison_to_literature}

We now consider the evolution of $\langle \beta \rangle$ with redshift across a larger redshift baseline by comparing to a variety of previous literature measurements.
The left-hand panel of Fig. \ref{fig:beta_vs_z_weighted_average} shows a comparison with other photometric analyses, while the right-hand panel shows a comparison to studies based primarily on spectroscopy. 
We crucially extend the redshift evolution of $\langle \beta \rangle$ down to $z=4$ using the \emph{HST} study of \citet{bouwens2014_uvslopes} (green diamonds).
To do this, we take the weighted average of the values in table 2 of \citet{bouwens2014_uvslopes} considering only $M_{\mathrm{UV}}$ values consistent with our data (i.e. $-21.25 < M_{\mathrm{UV}} < -17.25$).
It is immediately clear by-eye that these measurements, combined with our data, form a natural trend.
Motivated by the smooth evolution we see in comparison to the \citet{bouwens2014_uvslopes} measurements, and  in order to make a direct comparison with other literature studies, we fit a simple linear $\beta-z$ relation to our data (pink squares), finding
\begin{equation}
    \beta = (-0.11\pm0.02)\, z -(1.2\pm0.2).
    \label{eq:beta_z_combined_sample}
\end{equation} 
This relation is shown by the pink line in Fig. \ref{fig:beta_vs_z_weighted_average}.

It can be seen that extrapolating this relation to $z < 9$ yields a very good match to the \citet{bouwens2014_uvslopes} data.
If we fit a combined $\beta-z$ relation to our data and the \citet{bouwens2014_uvslopes} points, we find a fully consistent slope:
\begin{equation}
    \beta = (-0.097\pm0.004)\,z - (1.34\pm0.02).
\end{equation}
This relation is shown as a green line in Fig. \ref{fig:beta_vs_z_weighted_average}. 
We therefore find that, for galaxies in the range $-21.25 < M_{\mathrm{UV}} < -17.25$ the average UV continuum slope evolves smoothly to progressively redder values between $z\simeq14$ and $z\simeq4$ with a gradient of $\mathrm{d}\beta/\mathrm{d}z \simeq -0.1$.
Whether this smooth trend continues beyond the $\beta = -2.6$ limit --- or stabilises at this value --- remains unclear.
We also note that, while our $z > 9$ sample and the \citet{bouwens2014_uvslopes} $4 < z < 8$ sample cover the same range of $M_\mathrm{UV}$ (and the sample selections are subject to similar limiting magnitudes in the relevant selection bands for their respective redshift bins), this simple comparison does not account for the fact that the galaxies will be at different evolutionary stages at $z\simeq4$ and $z\simeq12$.

The redshift evolution we find is consistent with previous literature studies including \citet{Topping2024} who report ${\mathrm{d}\beta/\mathrm{d}z = -0.07}$ for ${8 < z < 12}$, \citet{Saxena2026} who report ${\mathrm{d}\beta/\mathrm{d}z = -0.08}$ for ${z\geq9.5}$ and \citet{Austin2025} who report ${\mathrm{d}\beta/\mathrm{d}z = -0.10}$ for ${6.5 \le z \le 13.5}$.
Moreover, it can be seen from the left panel of Fig. \ref{fig:beta_vs_z_weighted_average} that, at $z > 9$, the normalization of our $\beta-z$ relation is also consistent with the measurements of \citet{Hainline2026} and \citet{Topping2024} (as well as, unsurprisingly, our previous work; \citealp{Cullen2024}). 
At $z < 8$ the \citet{Topping2024} constraints fall below our best-fitting $\beta-z$ relation, but this discrepancy is readily explained by the fact that the \citet{Topping2024} sample is on average fainter ($M_{\mathrm{UV}} (\mathrm{median}) = -18.6$) in this redshift range than our sample ($M_{\mathrm{UV}} (\mathrm{median}) = -19.4$), and consequently fainter than the $M_{\mathrm{UV}}$-matched \citet{bouwens2014_uvslopes} data at the same redshifts.
Similarly, \citet{Austin2025} find bluer $\beta$ values on average at fixed $z$ (Fig. \ref{fig:beta_vs_z_weighted_average}).
However, the difference in this case cannot be easily explained by a difference in $M_\mathrm{UV}$ as the \cite{Austin2025} sample is actually slightly brighter than ours ($M_{\mathrm{UV}} (\mathrm{median}) = -19.8$) and should therefore be, if anything, redder.
The discrepancy in this case most likely stems from differences in sample selection and methodology.
However, it is worth noting that the difference is only in normalization, and that \citet{Austin2025} do find a ${\mathrm{d}\beta/\mathrm{d}z}$ slope consistent with our estimate.

We note that the gradient we find here is somewhat shallower than our original estimate of $\mathrm{d}\beta/\mathrm{d}z = -0.28$ based on the $9 < z < 12$ sample in \citet{Cullen2024}.
It can be seen from Fig. \ref{fig:beta_vs_z_weighted_average} that this was due to the narrow redshift range used in \citet{Cullen2024}.
Overall, across the numerous photometric analyses now available, there seems to be a growing consensus for a slope of $\mathrm{d}\beta/\mathrm{d}z \simeq -0.1$ in $M_{\mathrm{UV}}$-matched samples.

\subsubsection{Comparing with spectroscopic constraints}\label{sec:comparison_to_spectroscopy}

When looking at the right-hand panel of Fig. \ref{fig:beta_vs_z_weighted_average}, it can be seen that our results, and indeed our combined best-fitting $\beta-z$ relation, are generally in good agreement with the values reported in spectroscopic analyses, although with some exceptions.
The largest difference is seen with respect to \citet{robertsborsani2024_prismstacks}, who report UV slopes measured from stacked \emph{JWST}/NIRSpec PRISM spectra of star-forming galaxies at ${5 < z < 11}$ and report extremely blue average UV slopes at all redshifts, even down to $z\simeq6$.
Essentially all of these constraints fall below the photometric estimates of $\langle \beta \rangle$ at the same redshifts.
However, when taking weighted averages of the individual $\beta_\mathrm{spec}$ measurements from a more recent analysis presented in \citet{robertsborsani2026_morestacking}, we find better agreement. 
Comparing to these latest constraints, our photometric results are in decent agreement at $z\gtrsim12$, but are slightly bluer in the redshift range $10 < z < 12$.
Similarly, \citet{Saxena2026} find average spectroscopic $\beta$ values in line with our photometric estimate at $z \simeq 11$, but also find bluer values than our relation would predict at $z < 10$.
Finally, we find excellent consistency with the values measured in \citet{Tang2026} which fall along our best-fitting relation at $z=9-11$.

Although a significant variance remains between independent spectroscopic samples, overall it seems reasonable to conclude from Fig. \ref{fig:beta_vs_z_weighted_average} that the values derived from spectroscopic and photometric analyses are converging.
Indeed, given the small sample sizes and inherent incompleteness of spectroscopic datasets, it makes sense --- and is encouraging --- that as the sample size has grown these spectroscopic constraints are becoming increasingly more aligned with the larger and more complete photometric samples.
Nevertheless, it is clear that more work is needed in this area.
Ultimately, large and complete spectroscopic samples are required to definitively prove that the galaxy population reaches the intrinsic attenuation-free limit at $z > 10$, and to determine the intrinsic scatter of individual UV slopes at these early cosmic epochs.

\subsection{The $\mathbf{\beta-M_{\mathrm{UV}}}$ relation}\label{sec:muv_vs_beta_relation}

It is well established that at $z < 10$ there exists a relationship between absolute UV magnitude and $\beta$ (the so-called colour-magnitude, or $\beta-M_{\mathrm{UV}}$ relation; \citealp{bouwens2014_uvslopes}, \citealp{rogers2014_colormag}, \citealp{Cullen2024}, \citealp{Topping2024}). 
Due to sample sizes that lacked a sufficient dynamic range in $M_{\mathrm{UV}}$, studying the $\beta-M_{\mathrm{UV}}$ relation at $z>10$ initially proved difficult \citep{Cullen2024, Topping2024}.
However, the first constraints at these redshifts have recently been reported by \citet{Hainline2026}.
With our new sample of $\mathrm{N}=55$ galaxy candidates at $z\gtrsim 12$, covering an impressive range in $M_{\mathrm{UV}}$ (Fig. \ref{fig:MUV_vs_z}), we are in a position to provide an independent assessment at these redshifts.

\begin{figure}
    \centering
    \includegraphics[width=\linewidth]{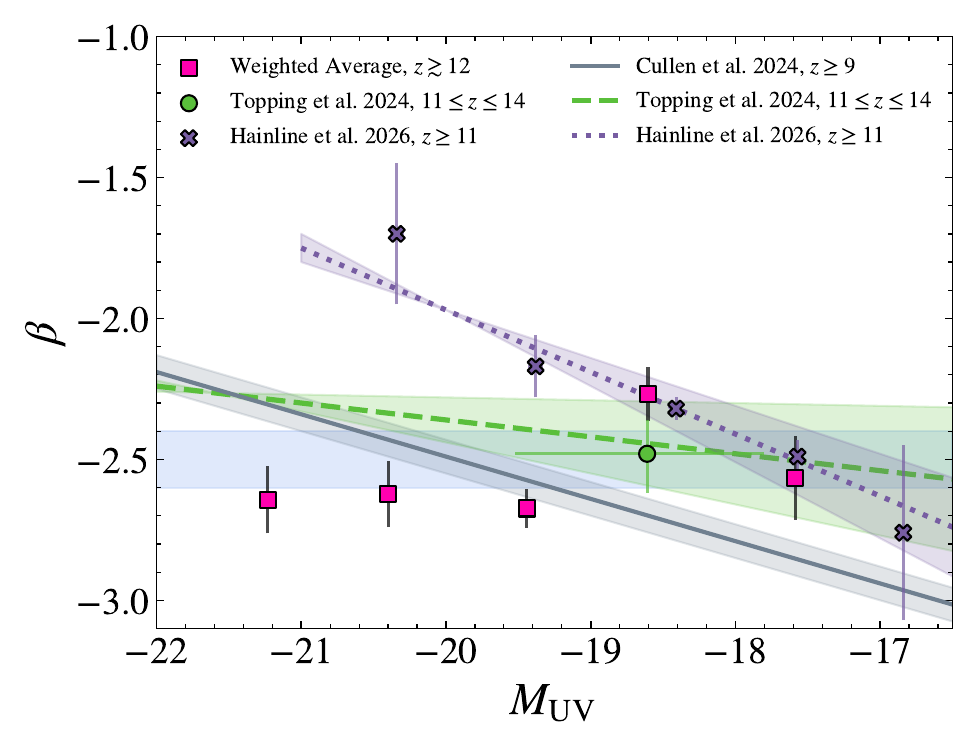}
    \caption{Plot of UV continuum slopes as a function of absolute UV magnitude ($M_{\mathrm{UV}}$). 
    The pink squares show the inverse variance-weighted average UV continuum slopes for our $z\gtrsim 12$ sample.
    We find no evidence for a $\beta-M_{\mathrm{UV}}$ relation in our sample: the average UV slope is consistent with the intrinsic limit (indicated by the horizontal blue band) at all magnitudes.
    Various literature relations are also shown, including \citet{Cullen2024} (grey line), \citet{Topping2024} (green dashed line), and \citet{Hainline2026} (purple dotted line).
    The redshift ranges for these various relations are indicated in the legend.
    The purple crosses show the average values from \citet{Hainline2026}.
    Although we see good agreement at the faint end, at the bright end there is a clear discrepancy between our data and the \citet{Hainline2026} $z\geq11$ sample.
    We discuss the possible reasons for this offset in Section \ref{sec:red_galaxies?}.
     }
    \label{fig:beta_VS_MUV_primarysample}
\end{figure}

\begin{table}
    \centering
    \caption{The inverse variance-weighted absolute UV magnitude $\langle M_{\mathrm{UV}} \rangle $ and UV continuum slope $\langle \beta \rangle $ for our $z\gtrsim12$ galaxy sample.}
    \begin{tabular}{|c|c|c|c|} 
    \hline
       $M_\mathrm{UV}$ Range & Number & $\langle M_\mathrm{UV} \rangle$& $\langle\beta\rangle$ \\
       \hline
         $M_\mathrm{UV}<-21$ &6 & $-21.23\pm0.03$  & $-2.64\pm0.12$ \\
         $-21\leq M_\mathrm{UV} < -20$ &7& $-20.40\pm0.04$ & $-2.62\pm0.12$   \\
         $-20\leq M_\mathrm{UV} < -19$ &27& $-19.44\pm0.02$   & $-2.67\pm0.07$  \\
         $-19\leq M_\mathrm{UV} < -18$ &10& $-18.60\pm0.03$ & $-2.27\pm0.10$  \\
         $-18\leq M_\mathrm{UV} < -17$ &5 &$ -17.59\pm0.03$ & $-2.57\pm0.15$   \\
        \hline 
    \end{tabular}
    \label{tab:weighted_averages_MUV}
\end{table}

Fig. \ref{fig:beta_VS_MUV_primarysample} shows the $\beta-M_{\mathrm{UV}}$ relation for our $z\gtrsim 12$ sample compared with a number of previous literature estimates.
As in Figs. \ref{fig:beta_vs_z} and \ref{fig:beta_vs_z_weighted_average}, we show the inverse variance-weighted mean values for our $z\gtrsim12$ sample as pink squares (values given in Table \ref{tab:weighted_averages_MUV}).
For clarity, we do not show the individual constraints in Fig. \ref{fig:beta_VS_MUV_primarysample} (see Fig. \ref{fig:beta_vs_z} for a useful indication of the typical scatter in the individual $\beta$ constraints).
We find that at all magnitudes $\langle \beta \rangle$ is close to the minimum intrinsic stellar plus nebular value.
In some sense this is unsurprising, given our general finding that by $z \gtrsim 11$ the typical galaxy is consistent with this attenuation-free limit.
We do not attempt to fit a relation to our data, as visually it can clearly be seen that these values are consistent with scatter about the sample-average value of $ \langle \beta \rangle = -2.57\pm0.04$.
This finding of a flat $\beta-M_{\mathrm{UV}}$ relation (i.e., no $\beta-M_{\mathrm{UV}}$ relation) is in clear contrast to the situation at lower redshifts, where a slope of $-0.2 < \mathrm{d}\beta/\mathrm{d}M_{\mathrm{UV}} < -0.1$ is commonly reported (see e.g. fig. 8 of \citealp{Cullen2024}). 
 
Comparing to other literature estimates, we find some interesting differences.
Focusing first on the analyses of \citet{Cullen2024} and \citet{Topping2024} we find a general consistency in the approximate normalisation, although both of these studies report $\beta-M_{\mathrm{UV}}$ gradients.
However, in both these cases---due to a lack of dynamic range in the data---these gradients are fixed to values measured at lower redshifts and only the normalisation is fitted (to illustrate this, the single data point from \citealp{Topping2024} is shown in Fig. \ref{fig:beta_VS_MUV_primarysample}).
In light of this we consider our results fully consistent with these early investigations. 

In contrast, \citet{Hainline2026} find ${\mathrm{d}\beta/\mathrm{d}M_\mathrm{UV} = -0.22\pm0.05}$ for their $z\geq11$ sample clearly contrasting with our findings.
Although our data are fully consistent with \citet{Hainline2026} at $M_\mathrm{UV} > -19$, there is a clear divergence at the bright end.
The reason for this discrepancy is due to significant differences in our sample selection techniques, specifically in the criteria applied to consider a high-redshift candidate robust.
We return to a detailed discussion of this issue in Section \ref{sec:discussion} below.

\section{Discussion}\label{sec:discussion}

The primary finding of this work is that at $z\gtrsim 12$ we find, on average, an extremely blue galaxy population with the typical UV slope at these redshifts being consistent with the intrinsic, attenuation-free, limit of $\beta \simeq -2.6$.
We find that the evolution of the typical UV slopes of galaxies with $-21 \lesssim M_{\mathrm{UV}} \lesssim -17$ is consistent with a smooth evolution to bluer values between $z\simeq4$ to $z\simeq15$ (Fig. \ref{fig:beta_vs_z_weighted_average}), with $z \gtrsim 11$ marking the epoch at which the average value has reached the intrinsic limit.
We also find a flat $\beta-M_{\mathrm{UV}}$ relation in our new $z\gtrsim 12$ sample indicating that--on average--all galaxies are attenuation-free regardless of their observed brightness.

In light of some recent results reporting individual galaxies with relatively red UV slopes $z \geq 10$ \citep[e.g.][]{mitsuhashi2026, donnan_2025ApJ_dust, Hainline2026, Rodighiero2026}, below we investigate whether there is any evidence to suggest that the blue UV slope values we find are a result of a bias in our selection criteria or measurement methodology.
We then discuss the existence of individual red sources within our $z > 12$ sample and consider whether (or not) these are consistent with the general evolutionary trend towards an attenuation-free population.

\subsection{Is our sample biased towards blue $\beta$ values?}\label{sec:simulation_section}

\begin{figure*}
    \centering
    \includegraphics[width=\linewidth]{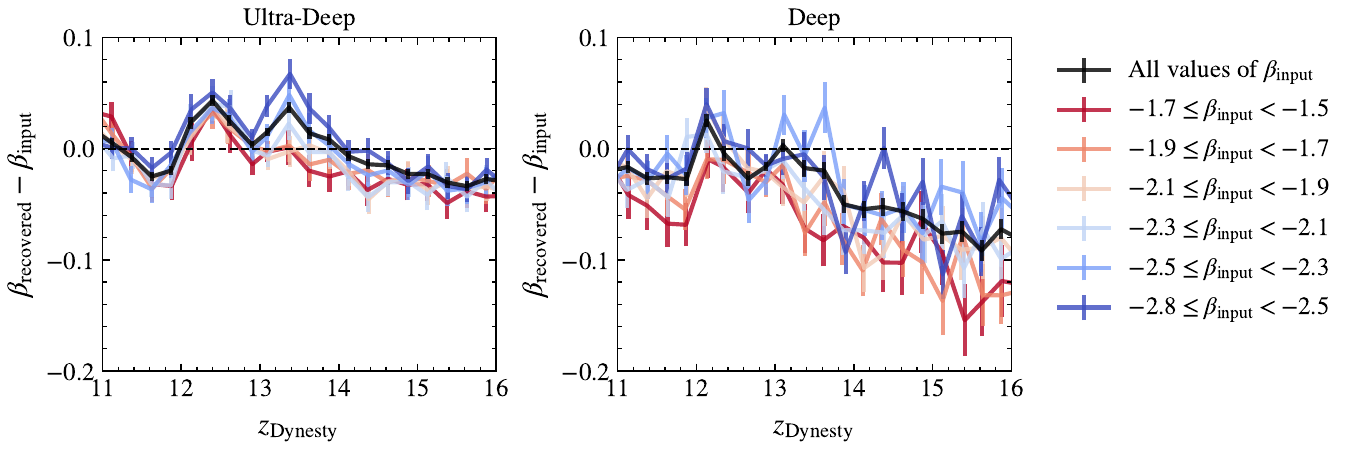}
    \caption{The difference between recovered $\beta$ and input $\beta$ for a suite of simulated galaxies as a function of their recovered redshifts.
    The panels showcase two depth tiers (see Section \ref{sec:simulation_section} and panel headings). 
    The black line shows the trend across all of the input $\beta$ values, and the coloured lines show the trend in bins of $\beta_\mathrm{input}$ as indicated in the legend. 
    For the deep tier, we find a small bias of $|\Delta \beta| \simeq 0.05$ that is only mildly dependent on the redshift and input $\beta$. This bias is even lower for the ultra-deep tier.}
    \label{fig:simulation_betadiff_beta_binned}
\end{figure*}

Because the average UV continuum slopes we measure are extremely blue, it is important to determine whether our sample could be inherently biased towards bluer sources (as a consequence of some combination of sample selection effects and measurement bias).  
To do this, we generate galaxy SEDs across a wide range of $\beta$ and produce simulated photometry of these objects assuming typical depths consistent with our observations.
We consider fixed $\beta$ values between $-1.5$ and $-2.8$ (with a step size of $0.1$), and for each $\beta$ value simulate $10 \, 200$ SEDs within the redshift range $10.0 \leq z \leq 17.0$ and UV magnitude range $-21.5\leq M_\mathrm{UV} \leq -17.0$. 
To reflect the varying depths of the surveys used to construct our sample, we run separate simulations for two different depth tiers: an ultra-deep tier, based on the depths in JADES GOODS-South region, and a deep tier, based on median depths of the CEERS and PRIMER fields (which are very similar to the median depths of the pure parallel fields included in this study). 
The exact depths used for these two tiers are calculated using table A1 in \citet{mcleod2026} and are used to perturb the model photometry and generate realistic observed data.
For each depth set, we use a filter-set consistent with the filters used in the representative programs. 
Overall, this results in simulated photometry for $\simeq 288 \, 000$ objects. 

To mimic our selection methodology, all of the simulated objects are fit with $\textsc{LePhare}$ to obtain photometric redshift estimates.
We then apply the selection criteria as described in Section \ref{sec:catalogue_construction} to retain only what we would consider robust high-redshift sources.
This process left $\simeq 145\, 000$ selected sources.
Assessing the $\beta$ distribution of the recovered sources, we find that our selection criteria do not preferentially select for bluer $\beta$ values.
The fraction of sources recovered is essentially flat as a function of input $\beta$, with a slight drop of $\simeq 10$ per cent at the very red end ($\beta > -1.7$).
This is unsurprising, since our selection criteria allow for a $\mathrm{SNR} > 8$ detection in \emph{either} of the $F200W$ or $F277W$ filters, which means we can in principle select objects with red slopes that are fainter at $\simeq 0.15 \, \mu\mathrm{m}$ and brighter at $\simeq 0.2 \, \mu\mathrm{m}$ (rest-frame).
Indeed, we reached a similar conclusion in our previous work (see appendix C of \citealp{Cullen2024}).
The slight drop-off at the reddest values of $\beta$ reflects the fact that these objects are more likely to be confused with low-redshift quiescent sources.
However, they are only marginally less likely to be selected: if they existed in large numbers, they would be present in our sample.

A separate question is whether our methodology for fitting $\beta$ results in a blue bias.
For the remaining $\simeq 145\, 000$ objects, we measure $\beta$ using the methodology described in Section \ref{sec:uvslope_fitting_photometry}.
In Fig. \ref{fig:simulation_betadiff_beta_binned} we show the difference between the input $\beta$ and the recovered $\beta$ (inverse variance-weighted) as a function of the best-fitting redshift obtained via our fitting method.
The panels show the two different depth tiers as indicated in their titles. 
The black line in each panel shows the running median value of this difference ($\Delta \beta$) across all values of input $\beta$, and the various coloured lines break this down into six $\beta$ ranges as indicated in the legend.
From Fig. \ref{fig:simulation_betadiff_beta_binned} it can be seen that, in general, we do see a small blue bias across all UV continuum slope values and redshifts, at least in the `Deep' tier.
This bias is consistently at the level of $-0.05 \lesssim\Delta \beta \lesssim 0.05$, with on average a slight, but essentially negligible bias blue (with a mild dependence on input $\beta$).
The fact that we find a slight blue bias in our photometric constraints is consistent with the small offset discussed with respect to the spectroscopic measurements in Fig. \ref{fig:beta_spec_vs_beta_phot_comparison}.
The bias does clearly worsen at $z>14$, although not at a level sufficient to explain $\langle \beta \rangle \simeq -3$ in our observed $z > 14$ candidates.
Overall however, the small bias we find here does not change any of the primary conclusions of this work.

As a final test, we investigate whether the bias is UV magnitude-dependent.
In the pre-\emph{JWST} era, it was common for fainter objects to be biased blue at the redshift-limit of \emph{HST} studies (e.g. \citealp{rogers2014_colormag}).
Separating our simulated sample into bright ($M_\mathrm{UV} \leq -20$) and a faint ($M_\mathrm{UV} > -20$) subsamples, we find no evidence for a larger blue bias in either, with the trend remaining the same as for the full sample. 
Again, this lack of $M_\mathrm{UV}$ dependence is unsurprising since the SEDs of all sources are well-sampled by the available photometry redward of the Lyman break, and our strict $\mathrm{SNR} > 8$ criteria ensures that all sources are well-detected regardless of their intrinsic magnitude.

\subsection{Evidence for red, dusty galaxies at $\mathbf{z\simeq12}$}\label{sec:red_galaxies?}

Recently, a number of works have reported instances of `red' galaxies at $z>10$, including sources with $\beta>-2$ (e.g. \citealt{mitsuhashi2026,donnan_2025ApJ_dust,Hainline2026, Rodighiero2026}).
It is important to consider these sources in the context of our claim that the typical UV slope of the galaxy population is consistent with the intrinsic attenuation-free limit at $z \gtrsim 11$.

Of these, the most robust is EGS-z11-R0, a galaxy at ${z_{\rm spec}=11.452 \pm 0.021}$ that has a NIRSpec/PRISM spectrum taken as part of \emph{JWST} GO program 4106 (PI: E. Nelson). 
\citet{Rodighiero2026} have recently presented an analysis of this source, reporting a remarkably red spectroscopic slope of $\beta_{\rm spec}=-0.68\pm0.3$.
This source was not selected in our initial \citet{Cullen2024} F115W-dropout sample as it fell on the edge of the short-wavelength imaging of our reduction.
The same edge effect is present in the v1.0 CEERS data release \citep{bagley2023,finkelstein2025}, and so the object does not feature in our new sample either.
From the publicly-available DJA spectrum, we find a somewhat bluer spectroscopic UV slope of $\beta_{\rm spec}=-1.51^{+0.15}_{-0.16}$, which is more consistent with the photometric measurement from \citet{Rodighiero2026}.
Despite the difference in our spectroscopic $\beta$ estimates, it nevertheless appears that this source is robustly $\beta > -2$.

However, it is clear from Fig. \ref{fig:beta_vs_z} that the existence of individual objects with such red $\beta$ values is perfectly consistent with our analysis.
For example, in our $z \gtrsim 12$ sample $\mathrm{N}=4/55$ ($\simeq 10$ per cent) have $\beta \gtrsim -2$.
Similarly red sources are also clearly present at $z > 10$ in the \citet{Cullen2024} sample.
Within the context of their `attenuation-free model', \citet{ferrara2026} interpret these redder sources as representing a `red monster' evolutionary phase, during which galaxies form stars at a moderate rate and build up ISM dust \citep[see also][]{Martinez-gonzales2026}.
This phase is followed by a powerful starburst that ultimately generates outflows to efficiently remove the dust and produces the more commonly-observed `blue monsters' (e.g. \citealt{Ziparo2023}).
One simple interpretation of the observations is therefore that the majority of $z > 10$ galaxies are undergoing an extreme starburst.
The quieter phase of dust build-up is either very transitory, or remains mostly hidden at the depth of \emph{JWST} observations.
The latter scenario seems unlikely as it would imply further boosting the surprisingly large abundance of $M_\mathrm{UV}$-bright galaxies at these redshifts \citep[e.g.][]{Ferrara2023, mcleod2024}.

Beyond the spectroscopic confirmation of the red $z\simeq11$ galaxy EGS-z11-R0, some authors have suggested that the inclusion of \emph{JWST}/NIRCam medium band photometry increases the likelihood of detecting redder sources at the highest redshifts (e.g., \citealt{asada2026}). 
The implication here is that earlier studies relying mostly on the \emph{JWST}/NIRCam wide bands \citep[e.g.][]{Topping2024, Cullen2024} are biased against red galaxies.
Recently \citet{mitsuhashi2026}, leveraging the additional medium band NIRCam filter coverage of Abell 2744 from the MegaScience survey (GO 4111; PI Suess), identified two red $z\simeq12$ candidates with $\beta=-1.33^{\,+0.09}_{-0.08}$ and $\beta=-1.54^{\,+0.07}_{-0.08}$. 
The authors specifically make the claim that there might be a bias towards selecting bluer galaxies when using broadband photometry only. 
However, both of these sources were selected in our previous broadband-selected samples with UV continuum slopes of $\beta = -1.49\pm0.25$ and $-1.65\pm0.25$ \citep{mcleod2024, Cullen2024}. 
As discussed in Section \ref{sec:simulation_section} above, we do not find any evidence for a bias towards preferentially recovering blue objects with our selection function (see also appendix C of \citealp{Cullen2024}).
The existence of objects with $\beta > -2$ in our current sample is also a testament to the fact that broadband selection can -- and does -- select red objects.
Nevertheless, the inclusion of medium bands is obviously beneficial in providing greater spectral sampling and thus tighter constraints on individual $\beta$ measurements.

Perhaps the most significant discrepancy between our results and results in the literature is with the $z \geq 11$ $\beta-M_{\mathrm{UV}}$ relation reported by \citet{Hainline2026} (see Fig. \ref{fig:beta_VS_MUV_primarysample}).
Specifically, while we find a flat relation (i.e. galaxies being consistent with the intrinsic dust-free limit at all $M_{\mathrm{UV}}$), \citet{Hainline2026} instead find a strong $\beta-M_{\mathrm{UV}}$ relation with the brightest sources (${M_{\mathrm{UV}} \lesssim -20}$) having very red UV slopes on average (${\langle \beta \rangle > -2}$).
To investigate this discrepancy we use the publicly-available photometry catalogue provided by \citet{Hainline2026}.
Focusing on the $2082$ sources at $z>8$, we first remove $\mathrm{N}=683/2082$ that do not meet our SNR selection criteria (see Section \ref{sec:catalogue_construction}).
We fit the remaining sources with \textsc{LePhare} and discard a further $\mathrm{N}=161/1399$ galaxies that either have a primary $z_\mathrm{phot} < 8$, $\chi^2_\mathrm{reduced} > 10$ or have a low redshift solution within $\Delta \chi^2=4$ of the high redshift solution.
This results in a final sample of $1238$ sources that mimics our selection criteria, highlighting the most obvious difference between our analyses: our stricter $8\sigma$ selection criteria means that we would only consider $\simeq 60$ per cent of the \citet{Hainline2026} sources to be robust $z>8$ galaxies.

Fitting $\beta$ and $M_{\mathrm{UV}}$ for the remaining $1238$ sources following the methodology described in Section \ref{sec:uvslope_fitting_photometry}, we find a flat $\beta-M_{\mathrm{UV}}$ relation that is much more consistent with the result we derive from our $z\gtrsim 12$ sample.
Of the three $z>11$ galaxies with $M_{\mathrm{UV}} < - 20$ in the \citet{Hainline2026} sample with $\langle \beta \rangle = -1.70 \pm 0.25$ (see Fig. \ref{fig:beta_VS_MUV_primarysample}), we find that only one source meets our selection criteria.
The retained source is the bluest of the three ($\beta = -2.33\pm0.05$ in \citealp{Hainline2026}; $\beta = -2.17^{+0.25}_{-0.25}$ following our methodology).
The removed sources are both very red ($\beta = -1.70\pm0.20$ and $\beta=-1.51\pm0.17$ in \citealp{Hainline2026}) and are responsible for the red average value.
One of these sources fails because of a statistically unacceptable fit ($\chi^2_{\nu} = 13$) and the other for having a lower redshift solution within $\Delta\chi^2 < 4$ of the $z > 11$ solution.
We show our version of the $\beta-M_{\mathrm{UV}}$ relation derived from the \citet{Hainline2026} sample in Appendix \ref{appendix:hainline_comparison}.
The comparison highlights how crucial the details of the sample selection can be in studies such as these.
Any impure selection is highly likely to bias samples to redder $\beta$, since the majority of interlopers whose photometry can mimic a $z > 9$ galaxy are by definition red sources.
Ultimately, we still await larger $z > 10$ spectroscopic samples to properly quantify the purity of the various selection techniques (e.g., \citealt{donnan2026b}).

To summarise, while our analysis is fully consistent with the existence of individual red and dust-reddened sources, the fraction of such objects is low (a crude estimate being $\simeq 10$ per cent with $\beta > -2$;  Fig. \ref{fig:beta_vs_z}).
Our main finding remains that \textit{the typical galaxy} at these redshifts has a UV slope between $\beta \simeq - 2.6$ and $\beta \simeq - 2.4$, consistent with the intrinsic attenuation-free limit.
We do not find any evidence to suggest that this is a result of a bias in our selection process.

\section{Conclusions}\label{sec:conclusions}

We have presented new measurements of the photometric UV continuum slopes $(\beta)$ of the $55$ galaxy candidates at $z \gtrsim 12$ from \citet{mcleod2026}. 
These galaxies were found through a search of $>0.6 \deg^2$ of \emph{JWST}/NIRCam imaging, spanning a redshift range of $11.5 \leq z \leq 15.2$ and an absolute UV magnitude range of $-21.6 \leq M_\mathrm{UV} \leq -17.3$ (Fig. \ref{fig:MUV_vs_z}). 
To increase the dynamic range in redshift, we supplemented the $z\gtrsim 12$ sample with our previous $z\ge9$ sample presented in \citet{Cullen2024}. 
We obtain UV continuum slopes by fitting a power-law of the form $f_\lambda \propto \lambda^\beta$ to the photometric data, and have validated our photometric measurements by comparing to available spectroscopic measurements (Fig. \ref{fig:beta_spec_vs_beta_phot_comparison}). 
In our analysis we focus mainly on the average values of the UV continuum slope $\langle\beta\rangle$ to characterize typical galaxy properties at $z \gtrsim 12$. Our main findings are: 
\begin{enumerate}

    \item We find a large range in individual $\beta$ measurements within the new $z \gtrsim 12$ sample ($-3.72 < \beta < -1.76$; Fig. \ref{fig:beta_vs_z}) but this is primarily a result of the significant uncertainties on the individual $\beta$ measurements (the median uncertainty is $\sigma_{\beta}\simeq0.3$). 
    Despite our sample containing many $\beta < -3$ sources, which would be potentially indicative of exotic stellar populations, we find no individual sources that are robustly this blue within the uncertainties.
    
    \item We do, however, find that on average the galaxy population at $z \gtrsim 12$ is extremely blue, with an inverse variance-weighted average of $\langle \beta\rangle = -2.57\pm0.04$.
    This value is consistent with the intrinsic $\beta$ limit for a stellar plus nebular continuum in the absence of dust-attenuation ($-2.6 \lesssim \beta \lesssim -2.4$). 
    Our results suggest that $z \gtrsim 12$ galaxies are consistent with being, on average, dust-attenuation free. 
    
    \item Our weighted averages at $z>10$ (Fig. \ref{fig:beta_vs_z} and Table \ref{tab:weighted_averages_beta}) are within the expected intrinsic $\beta$ range, forming a similar plateau of values as seen in \citet{Cullen2024}. 
    However, we find an evolution to redder values at $z\le10$ ($\langle\beta\rangle = -2.19\pm0.06$), suggesting that $z\simeq10-11$ marks the approximate epoch ($z_t$) at which galaxies become on average dust-attenuation free.
    Fitting two toy models to approximate $z_t$, we find $z_t \simeq 10.5$, consistent with the value reported in \citet{Cullen2024}.
    
    \item Fitting a linear $\beta-z$ relation to our data we find $\beta = (-0.11\pm0.02)z - (1.2\pm0.2)$ over the redshift range $9\le z \le 14$.
    Comparing this fit to an $M_{\mathrm{UV}}$-matched sample of galaxies down to $z\simeq4$ from \cite{bouwens2014_uvslopes}, we find that the extrapolation of this relation is in good agreement with the data at lower redshifts (Fig. \ref{fig:beta_vs_z_weighted_average}).
    Fitting to our data combined with \citet{bouwens2014_uvslopes}, so over the full range $4 < z < 14$, we find $\beta = (-0.097\pm0.004)\,z - (1.34\pm0.02)$.
    Our results suggest that the galaxy population with $-21 \lesssim M_\mathrm{UV} \lesssim -17$ becomes progressively bluer towards high redshifts with $\mathrm{d}\beta/\mathrm{d}z \simeq -0.1$.
        
    \item We find no evidence for a $\beta-M_\mathrm{UV}$ relation within our $z \gtrsim 12$ sample (Fig. \ref{fig:beta_VS_MUV_primarysample}).
    This result is in stark contrast to the situation at lower redshifts ($z \lesssim 10$), in which a clear $\beta-M_\mathrm{UV}$ relation is consistently observed with a gradient in the range $-0.2 < \mathrm{d}\beta/\mathrm{d}M_{\mathrm{UV}} < -0.1$.
    Instead, we find that the $z \gtrsim 12$ population is uniformly blue at all magnitudes and that the full galaxy population is therefore, on average, attenuation free.
    
    \item The lack of a $\beta-M_\mathrm{UV}$ relation in our sample contrasts with the results at $z \geq 11$ presented in \citet{Hainline2026}, who find that galaxies at the bright end ($M_\mathrm{UV} < -20$) are significantly reddened ($\beta \simeq -1.7$; Fig. \ref{fig:beta_VS_MUV_primarysample}).
    The core reason for this discrepancy is the different selection criteria between the two studies. 
    Specifically, when applying our selection cuts to the \citet{Hainline2026} sample, we find that only $\simeq 60 $ per cent of their sources are considered robust high-redshift sources based on our selection criteria.
    If we only consider these sources that would be recovered in our selection criteria, a similarly flat and blue $\beta-M_\mathrm{UV}$ relation is recovered (see discussion in Section \ref{sec:discussion} and Appendix \ref{appendix:hainline_comparison}).

    \item Running a suite of $\beta$ recovery simulations, we find that our sample selection criteria are not biased against redder galaxies, with the source recovery fraction dropping by only $\simeq10 $ per cent for $\beta=-1.7$.
    If bright dust-reddened sources with robust photometric redshifts at $z > 11$ existed they would be included in our sample.
    Moreover, we find that, at the typical depth of the surveys used, and for the filter sets available, the photometric $\beta$ measurements have only a small average bias of $|\Delta \beta| \simeq 0.05$.
    Applying a correction of this magnitude would not affect any of our main results.

    \item Finally, we discuss our results in the context of individual red sources with $\beta \gtrsim -2$ that have been discussed at $z \simeq 12$ \citep[e.g.][]{mitsuhashi2026, Rodighiero2026}.
    The distribution of individual $\beta$ measurements is fully consistent with the existence of individual red sources (Fig. \ref{fig:beta_vs_z}) although we find that the fraction of such sources is low ($\simeq 10$ per cent).
    Future follow up and confirmation of these early red sources could potentially provide crucial constraints on the nature of dust enrichment at the earliest cosmic epochs.
    
\end{enumerate}

In summary, our main result is that galaxies at $z\gtrsim12$ are \textit{on average} very blue, and consistent with the intrinsic dust-attenuation free $\beta$ limit. 
Overall, this is in good agreement with the trend discovered in early studies at slightly lower redshifts, and suggests that the galaxy population reaches --- and remains at --- the intrinsic $\beta$ limit at $z \gtrsim 10$ \citep[e.g.][]{Cullen2024, Topping2024, Austin2025}.
Although we have not considered them in detail here, there exist numerous theoretical explanations to explain the uniformly blue UV colours of these sources, including a lack of ISM dust either due to dust destruction \citep[][]{Esmerian2024}, strong outflows \citep[][]{Ferrara2023, Fiore2023, Ziparo2023}, a change in dust grain properties \citep[specifically a top-heavy grain distribution;][]{Narayanan_2026_growthofdust, Narayanan_2025ApJ, Sommovigo2026} which could result in a grey (i.e. flat) attenuation curve \citep{McKinney2026, Nakazato2026, Zimmerman2026}.
In order to further understand the physical mechanism that leads to the very blue galaxy population observed at $z\gtrsim 10$, we require continued spectroscopic follow-up of the current $z>10$ galaxy candidates with \emph{JWST} and, ultimately, improved constraints on their rest-frame IR emission, which will hopefully become possible with the upcoming \emph{ALMA} wide-band sensitivity upgrade.

\section*{Acknowledgements}
We want to thank everyone who has worked to enable the many \textit{JWST} surveys we use in this work. We also warmly thank Guido Roberts-Borsani for valuable insights and for providing data from \citet{robertsborsani2026_morestacking}.
DJM and JSD acknowledge the support of the Royal Society through the award of a Royal Society University Research Professorship to JSD.
FC acknowledges support from a UKRI Frontier Research Guarantee Grant (PI Cullen; grant reference: EP/X021025/1). ACC, H-HL, ET, and SDS acknowledge support from a UKRI Frontier Research Guarantee Grant (PI Carnall; grant reference EP/Y037065/1). Some of the data products presented herein were retrieved from the Dawn \textit{JWST} Archive (DJA). DJA is an initiative of the Cosmic Dawn Center (DAWN), which is funded by the Danish National Research Foundation under grant DNRF140.

This work is based primarily on observations made with the NASA/ESA/CSA {\it James Webb Space Telescope}. The data were obtained from the Mikulski Archive for Space Telescopes at the Space Telescope Science Institute, operated by the Association of Universities for Research in Astronomy, Inc., under NASA contract NAS 5-03127 for {\it JWST}. The observations used in this study are associated with programs 1063, 1176, 1180, 1181, 1199, 1208, 1210, 1283, 1286, 1287, 1324, 1345, 1355, 1433, 1727, 1810, 1837, 1840, 1895, 1963, 1967, 2079, 2234, 2282, 2514, 2555, 2561, 2566, 2727, 2732, 2736, 2738, 2739, 2750, 2756, 2767, 2883, 3073, 3215, 3293, 3362, 3516, 3538, 3859, 3990, 4043, 4111, 4125, 4212, 4287, 4598, 4625, 4744, 4446, 5224, 5398, 5893, 6368, 6434, 6511, 6541, 6549, 6556, 6564, 6882, 6954, 8060. The authors acknowledge the associated teams for developing their observing programs and in particular, we thank the programs that were developed and executed with a zero-exclusive-access period.

This work is also based (in part) on observations taken by the RELICS Treasury Program (GO 14096) with the NASA/ESA {\it Hubble Space Telescope}, which is operated by the Association of Universities for Research in Astronomy, Inc., under NASA contract NAS5-26555.

Some of the lensing clusters analysed come from the CLASH program with the NASA/ESA {\it Hubble Space Telescope} (GO 12065). The mass models, constructed in \citet{zitrin2009,zitrin2013,zitrin2015}, were obtained through the Hubble Space Telescope Archive, as a high-end science product of the CLASH program \citep{Postman2012}.

This work utilizes gravitational lensing models produced by PIs Bradač, Natarajan \& Kneib (CATS), Merten \& Zitrin, Sharon, Williams, Keeton, Bernstein and Diego, and the GLAFIC group. This lens modelling was partially funded by the HST Frontier Fields program conducted by STScI. STScI is operated by the Association of Universities for Research in Astronomy, Inc. under NASA contract NAS 5-26555. The lens models were obtained from the Mikulski Archive for Space Telescopes (MAST).

This research has made use of the SVO Filter Profile Service (http://svo2.cab.inta-csic.es/theory/fps/) supported from the Spanish MINECO through grant AYA2017-84089 \citep{Rodrigo2012, Rodrigo2020}.

This research has made use of NASA's Astrophysics Data System Bibliographic Services.

\section*{Data Availability}
At the time of writing, the photometric data sets used in this manuscript were all publicly available from the \textit{JWST} archive on MAST. For JADES and CEERS, these are available as part of their respective public data releases. All the spectra, except PAN-z14-1, are available on the DJA, with PAN-z14-1 becoming available in mid-November.

\bibliographystyle{mnras}
\bibliography{bibliography}

\newpage
\appendix

\section{Redshift evolution of the UV continuum slope with fixed redshift}
\begin{figure}
    \centering
    \includegraphics[width=\linewidth]{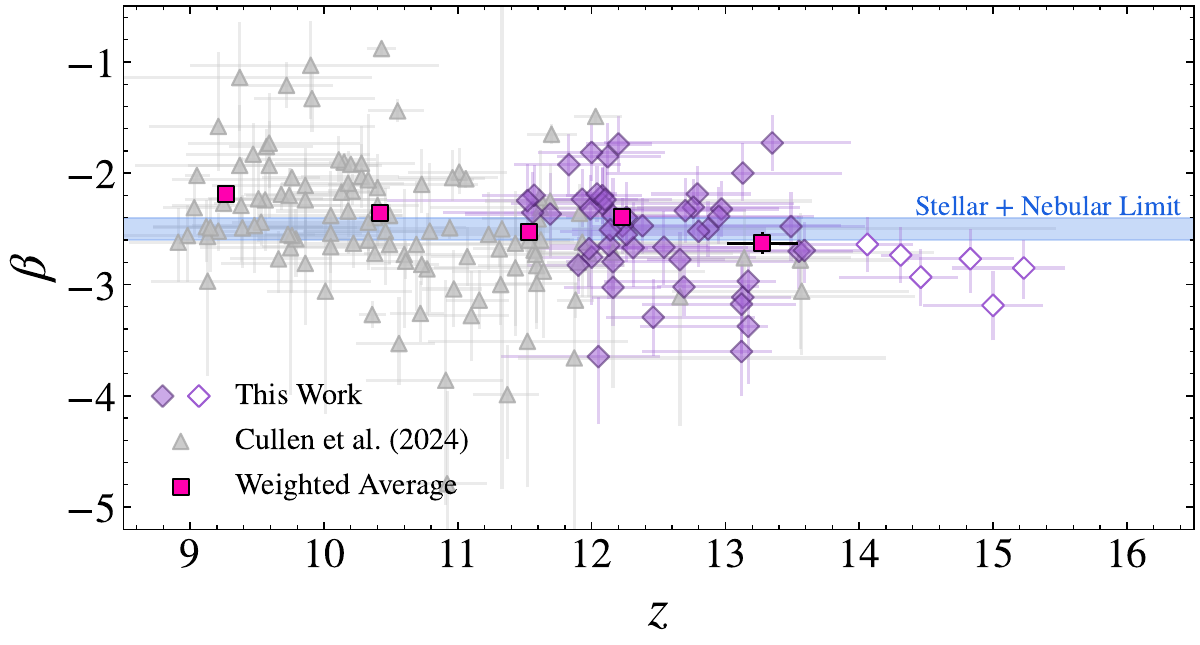}
    \caption{The same as Fig. \ref{fig:beta_vs_z} but with redshifts fixed to the redshifts obtained with the SED code in \citet{mclure2011}. The trend seen in Fig. \ref{fig:beta_vs_z} remains the same, with galaxies on average becoming blue ($\beta \sim -2.6$) at $z\gtrsim11$. }
    \label{fig:fixed_z_plot}
\end{figure} 
Here we investigate whether fixing the redshift to the photometric redshift obtained from fitting the full rest-frame UV to optical SED affects the $\beta-z$ relation. 
In Fig. \ref{fig:fixed_z_plot} it can be seen that the general trend remains as the trend we find in Fig. \ref{fig:beta_vs_z}, with the values of the UV continuum slope being consistent with the intrinsic limit at $z \gtrsim 11$.
The median differences between our adopted method (in which we use the photometric redshift as a prior rather than fixing it; see Section \ref{sec:uvslope_fitting_photometry}) and this fixed redshift approach is $\Delta\beta = -0.06$ and $\Delta z = 0.06$.

\section{The $\mathbf{M_\mathrm{UV}-\beta}$ relation of the Hainline et al. sample}\label{appendix:hainline_comparison}

\begin{figure}
    \centering
    \includegraphics[width=\linewidth]{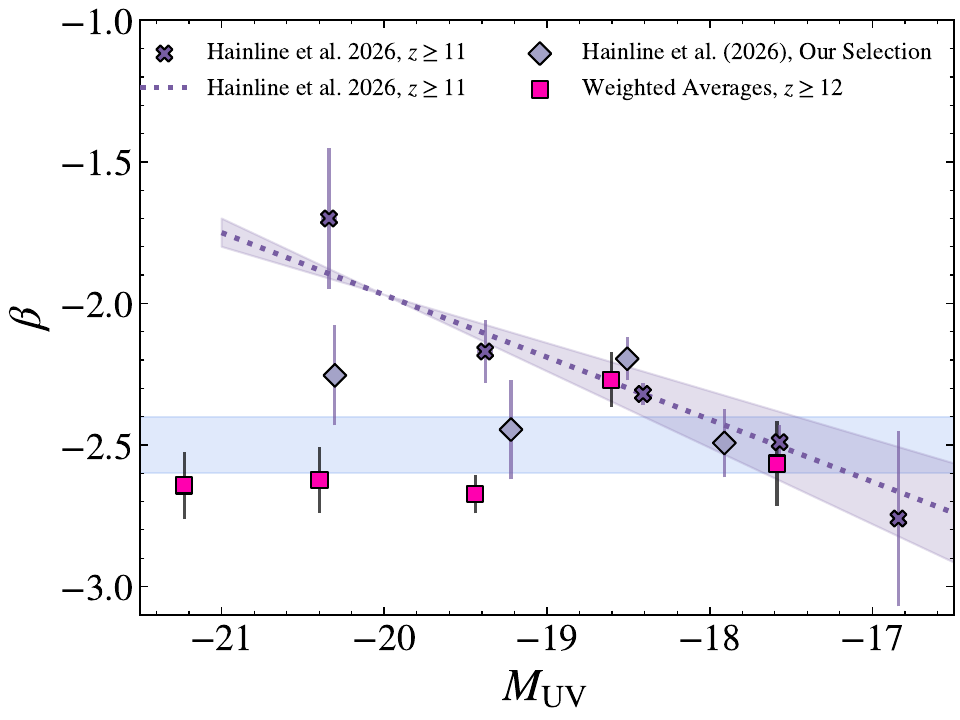}
    \caption{A plot showing the effect of applying our selection criteria to the $M_\mathrm{UV}-\beta$ relation of \citet{Hainline2026}.
    The purple crosses show the original values reported in \citet{Hainline2026} while the light purple diamonds show how these values change once our selection criteria are applied to their sample (which removes $\simeq 40$ per cent of their sources; see Section \ref{sec:red_galaxies?} for a full description).
    It is clear that, when considering only the sources that we would consider robust high-redshift candidates, the $M_\mathrm{UV}-\beta$ relation becomes essentially flat and much more consistent with our finding (pink squares).}
    \label{fig:MUV_beta_hainline_comp}
\end{figure}

Here we show the impact on the $M_\mathrm{UV}-\beta$ relation derived from the \citet{Hainline2026} sample if we apply our selection criteria (which removes $\simeq 40$ per cent of their candidates; see Section \ref{sec:red_galaxies?}).
In Fig. \ref{fig:MUV_beta_hainline_comp} it can be seen that applying our selection criteria results in an $M_\mathrm{UV}-\beta$ relation much more consistent with the one presented in this paper (i.e. consistent with a constant value of $\beta$ at all $M_\mathrm{UV}$).
The difference between the relation derived after our selection criteria are applied, compared to the relation presented in \citet{Hainline2026}, is especially stark at the bright end. 
We find that $\langle \beta \rangle$ in the ${-21 < M_\mathrm{UV} < -20}$ and ${-20 < M_\mathrm{UV} < -19}$ bins shifts from ${\langle \beta \rangle=-1.7\pm0.25}$ and ${\langle \beta \rangle=-2.17\pm0.11}$ (as reported in \citealp{Hainline2026}) to ${\langle \beta \rangle=-2.25\pm0.18}$ and ${\langle \beta \rangle=-2.45\pm0.18}$, respectively. 
\section{Tabulated $\beta$ for the full sample}

In Table \ref{tab:appendix_all_sources} we present the properties for the individual galaxies in our sample.

\begin{table*}\label{tab:appendix_all_sources}
    \centering
    \caption{The properties of the individual sources in our $z>12$ sample. The redshift ($z_{\rm Dynesty}$) and UV continuum slope ($\beta$) presented are from our fit with \texttt{Dynesty}. The absolute UV magnitude ($M_{\rm UV}$) is calculated at 1500\AA~and is corrected to total and corrected for de-lensing (using magnification $\mu$). Where applicable we also include the spectroscopic redshift ($z_{\rm spec}$) and UV continuum slope ($\beta_{\rm spec}$). Note that for objects with $\mu = 1.2$ this implies that the candidate is in a lensed field, but is outside the lensing map coverage. }
    \begin{tabular}{lccccccccl}
    \hline
        ID & RA & Dec & $z_\mathrm{Dynesty}$ & $\beta$ & $M_\mathrm{UV}$ & $\mu$ & $z_{\rm spec}$ & $\beta_{\rm spec}$ & Ref. \\ [1ex] \hline 
        2236+0032-2027652 & 22:36:44.57 & $+$00:32:18.57 & $11.5_{-0.8}^{+0.9}$ & $-2.25_{-0.39}^{+0.35}$ & $-20.0_{-0.1}^{+0.1}$ & - & -& - & - \\ [1ex]
        PAN-1707+5852-2015557 & 17:07:15.91 & $+$58:51:44.10 & $11.6_{-0.4}^{+0.4}$ & $-2.35_{-0.34}^{+0.32}$ & $-20.4_{-0.2}^{+0.2}$ & - & -& - & - \\ [1ex]
        PAN-0134-1531-2023208 & 01:34:43.79 & $-$15:31:27.39 & $11.6_{-0.5}^{+0.7}$ & $-2.35_{-0.4}^{+0.38}$ & $-20.5_{-0.2}^{+0.2}$ & - & -& - & - \\ [1ex]
        PAN-0104-5508-2026296 & 01:04:17.50 & $-$55:07:35.55 & $11.7_{-0.4}^{+0.4}$ & $-2.23_{-0.27}^{+0.28}$ & $-20.6_{-0.1}^{+0.1}$ & - & - & -& - \\ [1ex]
        UNCOVER-2079090 & 00:14:29.18 & $-$30:27:22.84 & $11.8_{-0.3}^{+0.4}$ & $-1.94_{-0.28}^{+0.25}$ & $-19.2_{-0.1}^{+0.1}$ & 1.2 & - & -& $\mu$: \citet{Furtak2023} \\ [1ex]
        PAN-1007+2109-2040861 & 10:07:44.02 & $+$21:11:00.98 & $11.9_{-0.3}^{+0.3}$ & $-2.83_{-0.28}^{+0.28}$ & $-20.7_{-0.1}^{+0.1}$ & - & - & -& - \\ [1ex]
        JADES-GS-2199196 & 03:32:39.92 & $-$27:49:17.62 & $12.0_{-0.1}^{+0.1}$ & $-2.22_{-0.25}^{+0.25}$ & $-19.0_{-0.1}^{+0.1}$ & - & 12.63 &  $-2.07^{+0.09}_{-0.09}$ & \citet{Curtis-Lake_2023} \\ [1ex]
        JADES-GS-2037358 & 03:32:06.88 & $-$27:53:34.85 & $12.1_{-0.2}^{+0.3}$ & $-2.23_{-0.25}^{+0.25}$ & $-18.5_{-0.1}^{+0.1}$ &- & - & - & - \\ [1ex]
        JADES-GS-2104976 & 03:32:25.12 & $-$27:51:42.74 & $12.1_{-0.3}^{+0.8}$ & $-2.38_{-0.35}^{+0.32}$ & $-18.0_{-0.1}^{+0.1}$ &- & - & - & - \\ [1ex]
        CEERS-2274136 & 14:19:12.35 & $+$52:51:41.81 & $12.1_{-0.4}^{+0.6}$ & $-3.62_{-0.7}^{+0.57}$ & $-18.9_{-0.0}^{+0.1}$ & - &- & - & - \\ [1ex]
        PAN-1343+5549-2000350 & 13:43:39.91 & $+$55:46:45.76 & $12.1_{-0.4}^{+1.1}$ & $-2.77_{-0.36}^{+0.35}$ & $-21.3_{-0.1}^{+0.1}$ &- & - & - & - \\ [1ex]
        PAN-1707+5852-2010068 & 17:07:17.36 & $+$58:51:11.06 & $12.1_{-0.4}^{+1.3}$ & $-2.28_{-0.3}^{+0.28}$ & $-21.6_{-0.2}^{+0.1}$ & - & - & -&-  \\ [1ex]
        GLASS-2039052 & 00:13:59.75 & $-$30:19:29.14 & $12.2_{-0.1}^{+0.0}$ & $-2.53_{-0.25}^{+0.25}$ & $-21.3_{-0.0}^{+0.0}$ & 1.2 & 12.34 & $-2.78^{+0.06}_{-0.06}$ & \citet{Castellano2024}, \\ 
        & & & & & & & & & $\mu$: \citet{Furtak2023}\\
        JADES-GN-3145713 & 12:37:30.39 & $+$62:15:15.68 & $12.2_{-0.2}^{+0.7}$ & $-2.57_{-0.32}^{+0.28}$ & $-19.9_{-0.1}^{+0.1}$ & - & - & -&-  \\ [1ex]
        CEERS-East-1160411 & 14:20:03.15 & $+$52:52:30.49 & $12.2_{-0.4}^{+0.7}$ & $-2.41_{-0.32}^{+0.32}$ & $-19.2_{-0.1}^{+0.1}$ & - & - & -&-  \\ [1ex]
        CEERS-2264171 & 14:19:27.31 & $+$52:51:29.20 & $12.3_{-0.4}^{+0.8}$ & $-2.28_{-0.35}^{+0.33}$ & $-19.3_{-0.1}^{+0.1}$ & - & - & -&-  \\ [1ex]
        BEA-0217-0503-2052739 & 02:17:13.11 & $-$05:01:50.54 & $12.3_{-0.4}^{+0.8}$ & $-2.77_{-0.56}^{+0.44}$ & $-19.0_{-0.1}^{+0.1}$ & - & - & - &- \\ [1ex]
        CEERS-East-1088018 & 14:20:12.28 & $+$52:48:32.66 & $12.3_{-0.5}^{+0.9}$ & $-2.85_{-0.48}^{+0.4}$ & $-18.4_{-0.1}^{+0.1}$ & - & - & -&-  \\ [1ex]
        PAN-1237+6210-1016125 & 12:37:46.16 & $+$62:11:49.72 & $12.3_{-0.5}^{+1.4}$ & $-1.96_{-0.39}^{+0.35}$ & $-19.5_{-0.3}^{+0.2}$ & - & - & -&-  \\ [1ex]
        J1235-2051316 & 12:35:53.39 & $+$04:55:53.04 & $12.3_{-0.6}^{+1.1}$ & $-2.36_{-0.37}^{+0.42}$ & $-18.8_{-0.2}^{+0.2}$ & - & - & - &- \\ [1ex]
        PAN-0144+1714-2042105 & 01:44:22.44 & $+$17:15:31.36 & $12.4_{-0.3}^{+0.7}$ & $-3.05_{-0.36}^{+0.28}$ & $-19.1_{-0.1}^{+0.1}$ & - & - & -&-  \\ [1ex]
        PAN-1000+0207-2032751 & 10:00:24.17 & $+$02:08:31.01 & $12.5_{-0.5}^{+0.9}$ & $-2.48_{-0.4}^{+0.35}$ & $-20.1_{-0.1}^{+0.1}$ & - & - & -&-  \\ [1ex]
        CEERS-East-3024423 & 14:19:49.00 & $+$52:44:06.57 & $12.5_{-0.5}^{+1.4}$ & $-1.89_{-0.45}^{+0.25}$ & $-19.5_{-0.4}^{+0.1}$ & - & - & -&-  \\ [1ex]
        NGDEEPMIDIS-2043718 & 03:32:58.43 & $-$27:51:02.90 & $12.6_{-0.4}^{+0.4}$ & $-2.72_{-0.41}^{+0.32}$ & $-17.6_{-0.1}^{+0.1}$ & - & - & - &- \\ [1ex]
        SPT0615-1020236 & 06:15:41.93 & $-$57:43:59.79 & $12.6_{-0.5}^{+0.5}$ & $-2.56_{-0.33}^{+0.32}$ & $-19.8_{-0.1}^{+0.1}$ & 1.2 & - & -& $\mu$: \citet{PaternoMahler2018}, \\
        & & & & & & & & & \phantom{oo}\,\citet{zitrin2015}\\

        BEA-2303-6250-2039802 & 23:03:34.42 & $-$62:50:11.26 & $12.6_{-0.5}^{+0.7}$ & $-2.61_{-0.25}^{+0.25}$ & $-21.1_{-0.1}^{+0.1}$ & - & - & - &- \\ [1ex]
        PRIMER-UDS-2564396 & 02:17:10.08 & $-$05:10:26.89 & $12.7_{-0.2}^{+0.3}$ & $-2.85_{-0.25}^{+0.26}$ & $-20.1_{-0.0}^{+0.1}$ & - & - & -&-  \\ [1ex]
        NGDEEPMIDIS-2051940 & 03:33:02.18 & $-$27:50:20.12 & $12.7_{-0.4}^{+0.3}$ & $-2.39_{-0.33}^{+0.33}$ & $-17.3_{-0.0}^{+0.1}$ & - & - & - &- \\ [1ex]
        NGDEEPMIDIS-2051280 & 03:33:00.74 & $-$27:50:21.41 & $12.7_{-0.5}^{+0.4}$ & $-2.35_{-0.25}^{+0.25}$ & $-18.0_{-0.1}^{+0.1}$ & - & - & - &- \\ [1ex]
        NGDEEPMIDIS-2041809 & 03:32:56.97 & $-$27:51:10.65 & $12.8_{-0.3}^{+0.3}$ & $-2.27_{-0.26}^{+0.25}$ & $-17.5_{-0.1}^{+0.1}$ & - & - & - &- \\ [1ex]
        MACS0416-1003914 & 04:16:16.35 & $-$24:07:41.72 & $12.8_{-0.4}^{+0.4}$ & $-2.59_{-0.25}^{+0.25}$ & $-19.3_{-0.1}^{+0.1}$ & - & - & - &- \\ [1ex]
        BEA-2303-6250-1006779 & 23:03:57.79 & $-$62:52:17.48 & $12.8_{-0.6}^{+0.8}$ & $-2.79_{-0.49}^{+0.47}$ & $-19.3_{-0.2}^{+0.2}$ & - & - & - &- \\ [1ex]
        CEERS-East-3092309 & 14:20:06.05 & $+$52:48:15.18 & $12.9_{-0.3}^{+0.3}$ & $-3.12_{-0.28}^{+0.25}$ & $-19.3_{-0.0}^{+0.0}$ & - & - & - &- \\ [1ex]
        JADES-GS-2042986 & 03:32:15.54 & $-$27:53:24.86 & $12.9_{-0.4}^{+0.4}$ & $-2.37_{-0.25}^{+0.25}$ & $-18.4_{-0.1}^{+0.1}$ & - & 13.01 &$-3.31^{+0.45}_{-0.58}$ & \citet{witstok_2025_z13_lya} \\ [1ex]
        NGDEEPMIDIS-2048620 & 03:32:58.20 & $-$27:50:38.72 & $12.9_{-0.5}^{+0.4}$ & $-3.43_{-0.37}^{+0.37}$ & $-17.7_{-0.0}^{+0.0}$ & - & - & - &- \\ [1ex]
        UNCOVER-2192918 & 00:14:12.63 & $-$30:21:00.76 & $12.9_{-0.5}^{+0.5}$ & $-2.45_{-0.31}^{+0.28}$ & $-18.7_{-0.1}^{+0.1}$ & 1.9 & - & - & $\mu$: \citet{Furtak2023} \\ [1ex]
        BEA-0446-2636-2034805 & 04:46:42.76 & $-$26:34:42.75 & $12.9_{-0.6}^{+0.5}$ & $-2.05_{-0.29}^{+0.28}$ & $-19.9_{-0.2}^{+0.1}$ & - & - & - & -  \\ [1ex]
        NEP-TDF-2034945 & 17:22:57.48 & $+$65:46:29.09 & $12.9_{-0.8}^{+0.7}$ & $-2.93_{-0.33}^{+0.26}$ & $-19.6_{-0.1}^{+0.1}$ & - & - & -&-  \\ [1ex]
        PRIMER-UDS-2138854 & 02:17:01.80 & $-$05:17:18.67 & $13.0_{-0.4}^{+0.4}$ & $-3.72_{-0.48}^{+0.4}$ & $-19.8_{-0.0}^{+0.0}$ & - & - & - &- \\ [1ex]
        BEA-2303-6250-2027277 & 23:03:47.47 & $-$62:51:07.34 & $13.0_{-0.6}^{+0.5}$ & $-3.22_{-0.37}^{+0.32}$ & $-20.0_{-0.1}^{+0.1}$ & - & - & -&-  \\ [1ex]
        JADES-GS-2084600 & 03:32:11.05 & $-$27:52:10.66 & $13.0_{-0.7}^{+0.6}$ & $-1.76_{-0.25}^{+0.25}$ & $-18.8_{-0.2}^{+0.1}$ & - & - & - &- \\ [1ex]
        BEA-2058-4247-2067175 & 20:58:33.74 & $-$42:44:08.61 & $13.1_{-0.4}^{+0.5}$ & $-3.31_{-0.5}^{+0.48}$ & $-21.2_{-0.1}^{+0.1}$ & - & - & -&-  \\ [1ex]
        PLCKG165-2047505 & 11:27:13.98 & $+$42:28:56.16 & $13.1_{-0.6}^{+0.4}$ & $-3.03_{-0.3}^{+0.25}$ & $-20.0_{-0.1}^{+0.1}$ & 3.2 & - & -& $\mu$: \citet{Kamieneski2024} \\ [1ex]
        MACS0416-1003912 & 04:16:16.33 & $-$24:07:41.72 & $13.2_{-0.5}^{+0.3}$ & $-3.56_{-0.58}^{+0.5}$ & $-19.0_{-0.1}^{+0.1}$ & - & - & -&-  \\ [1ex]
        \hline

            \end{tabular}
\end{table*}

\begin{table*}
    
    \centering
    \begin{tabular}{lccccccccl}
    \hline
        ID & RA & Dec & $z_\mathrm{Dynesty}$ & $\beta$ & $M_\mathrm{UV}$ & $\mu$ & $z_{\rm spec}$ & $\beta_{\rm spec}$&  Ref. \\ [1ex] \hline 
        JADES-GS-1074873 & 03:32:05.31 & $-$27:51:26.05 & $13.2_{-0.8}^{+0.6}$ & $-2.5_{-0.39}^{+0.31}$ & $-19.4_{-0.2}^{+0.1}$ & - & - & -&-  \\ [1ex]
        GLASS-2014204 & 00:14:03.26 & $-$30:21:24.40 & $13.3_{-0.8}^{+0.5}$ & $-2.68_{-0.34}^{+0.28}$ & $-19.9_{-0.1}^{+0.1}$ & 1.7 & - & -& $\mu$: \citet{Furtak2023} \\ [1ex]
        JADES-GS-1061386 & 03:32:20.33 & $-$27:52:00.02 & $13.3_{-1.3}^{+0.7}$ & $-2.05_{-0.4}^{+0.37}$ & $-18.2_{-0.3}^{+0.3}$ & - & - & -&-  \\ [1ex]
        CEERS-2507584 & 14:20:17.83 & $+$52:57:04.12 & $13.4_{-0.7}^{+0.5}$ & $-2.72_{-0.41}^{+0.34}$ & $-19.2_{-0.2}^{+0.1}$ & - & - & -&-  \\ [1ex]
        J1235-2062314 & 12:35:55.12 & $+$04:56:51.50 & $13.4_{-0.8}^{+0.5}$ & $-2.78_{-0.4}^{+0.34}$ & $-19.4_{-0.2}^{+0.1}$ & - & - & -&-  \\ [1ex]
        JADES-GS-2051088 & 03:32:17.83 & $-$27:53:09.30 & $13.9_{-0.6}^{+0.3}$ & $-2.62_{-0.38}^{+0.36}$ & $-19.1_{-0.1}^{+0.2}$ & - & 13.90 &$-2.95^{+0.18}_{-0.19}$& \citet{Carniani2024} \\ [1ex]
        PAN-2216+0024-3003658 & 22:17:00.12 & $+$00:22:44.69 & $14.3_{-0.2}^{+0.1}$ & $-2.82_{-0.25}^{+0.25}$ & $-21.2_{-0.1}^{+0.1}$ & - & 13.53 &$-2.29^{+0.1}_{-0.1}$ & \citet{donnan2026} \\ [1ex]
        PRIMER-COS-2332181 & 10:00:22.40 & $+$02:16:23.35 & $14.5_{-0.3}^{+0.3}$ & $-3.08_{-0.33}^{+0.31}$ & $-20.0_{-0.1}^{+0.1}$ & - & 14.44 &$-2.50^{+0.19}_{-0.18}$ & \citet{Naidu2026} \\ [1ex]
        JADES-GN-2285047 & 12:36:40.16 & $+$62:18:36.94 & $14.8_{-0.3}^{+0.2}$ & $-2.9_{-0.34}^{+0.3}$ & $-20.0_{-0.1}^{+0.1}$ & - & - &- & - \\ [1ex]
        PAN-2345-4235-3022235 & 23:45:15.45 & $-$42:34:57.07 & $15.2_{-0.3}^{+0.2}$ & $-2.94_{-0.29}^{+0.28}$ & $-20.9_{-0.1}^{+0.1}$ & - & - &- & - \\ [1ex]
        PAN-0332-2754-2030538 & 03:32:41.21 & $-$27:53:45.00 & $15.2_{-0.3}^{+0.3}$ & $-3.44_{-0.37}^{+0.38}$ & $-19.2_{-0.1}^{+0.1}$ & - & - &- & - \\ [1ex]
        \hline

    \end{tabular}
\end{table*}

\bsp	%
\label{lastpage}
\end{document}